\documentclass[journal]{IEEEtran}
\usepackage{ifpdf}
\usepackage{cite}
\ifCLASSINFOpdf
  \usepackage[pdftex]{graphicx}
\else
\fi
\usepackage{amsmath,amssymb,amsfonts}

\usepackage{latexsym}
\usepackage{multirow}

\usepackage{nomencl}
\makenomenclature
\usepackage{etoolbox}
\renewcommand\nomgroup[1]{%
  \item[\bfseries
  \ifstrequal{#1}{A}{Acronym}{%
  \ifstrequal{#1}{B}{Symbols}{}}%
]}

\usepackage[linesnumbered,ruled]{algorithm2e}

\makeatletter
\newcommand{\numberthis}{\refstepcounter{equation}\tag{\theequation}}
\makeatother

\SetKwInput{KwInput}{Input}                
\SetKwInput{KwOutput}{Output}              
\SetKwInput{KwIteration}{Iteration}  
\SetKwInput{KwInitialization}{Initialization}  

\usepackage{array}
\usepackage{fixltx2e}
\usepackage{stfloats}
\usepackage{caption}
\usepackage{subfigure}
\usepackage{xcolor}
\usepackage[switch]{lineno}

\begin{document}

\title{Graph Neural Network Aided Detection for the Multi-User  Multi-Dimensional Index Modulated Uplink}

\author{Xinyu~Feng,~\IEEEmembership{Member,~IEEE,}
        Mohammed EL-Hajjar,~\IEEEmembership{Senior Member,~IEEE,}
        Chao Xu,~\IEEEmembership{Senior Member,~IEEE,}
        and~Lajos~Hanzo,~\IEEEmembership{Life Fellow,~IEEE}}
\maketitle

\begin{abstract}
The concept of Compressed Sensing-aided Space-Frequency Index Modulation (CS-SFIM) is conceived for the Large-Scale Multi-User Multiple-Input Multiple-Output Uplink (LS-MU-MIMO-UL) of Next-Generation (NG) networks. Explicitly, in CS-SFIM, the information bits are mapped to both spatial- and frequency-domain indices, where we treat the activation patterns of the transmit antennas and of the subcarriers separately. Serving a large number of users in an MU-MIMO-UL system leads to substantial Multi-User Interference (MUI).  Hence, we design the Space-Frequency (SF) domain matrix as a joint factor graph, where the Approximate Message Passing (AMP) and Expectation Propagation (EP) based MU detectors can be utilized. In the LS-MU-MIMO-UL scenario considered, the proposed system uses optimal Maximum Likelihood (ML) and Minimum Mean Square Error (MMSE) detectors as benchmarks for comparison with the proposed MP-based detectors. These MP-based detectors significantly reduce the detection complexity compared to ML detection, making the design eminently suitable for LS-MU scenarios. To further reduce the detection complexity and improve the detection performance, we propose a pair of Graph Neural Network (GNN) based detectors, which rely on the orthogonal AMP (OAMP) and on the EP algorithm, which we refer to as the GNN-AMP and GEPNet detectors, respectively. The GEPNet detector maximizes the detection performance, while the GNN-AMP detector strikes a performance versus complexity trade-off.  The GNN is trained for a single system configuration and yet it can be used for any number of users in the system. The simulation results show that the GNN-based detector approaches the ML performance in various configurations.
\end{abstract}



\IEEEpeerreviewmaketitle

\section{INTRODUCTION}
\IEEEPARstart{M}{ultiple}-Input Multiple-Output (MIMO) schemes have been widely used in wireless systems for improving performance and Spectral Efficiency (SE). They have also been combined with Orthogonal Frequency Division Multiplexing (OFDM) for communication over frequency-selective channels~\cite{mutut}. Multi-User (MU) MIMO systems, which benefit from further improved SE~\cite{mutut} and Power Efficiency {(PE)}~\cite{MIMOdet}, are widely considered as a key technique for Next-Generation (NG) wireless systems~\cite{massivemimo1}.  However, a persistent challenge in MU-MIMO systems is the presence of Multi-User Interference (MUI)~\cite{massivemimo2}. \par
Index Modulation {(IM)} \cite{im1} constitutes a potential candidate for next-generation wireless systems as a benefit of its flexible resource activation~\cite{smhistory}. The concept of {IM} has been derived from that of Spatial Modulation {(SM)}~\cite{sm1}\cite{chaosm}, which is capable of striking a flexible performance  versus complexity trade-off  using a single Radio Frequency {(RF)} chain~\cite{massivesm,smdesign} along with low-complexity detection.  To achieve improved flexibility, generalized SM (GSM) has been proposed, where multiple Transmit Antennas {(TA)} can be activated simultaneously~\cite{gsm} and utilized for MU scenarios~\cite{gsmmu}.\par

Then, the concept of {SM} has been extended to the frequency domain, where the resultant philosophy of {IM} has been proposed in~\cite{imconcept}.  OFDM associated with index modulation {(OFDM-IM)}, which has also been referred to as Subcarrier-{IM} {(SIM)}~\cite{sim} was proposed for increasing the Energy Efficiency (EE) and performance compared to traditional OFDM, where only a fraction  of the subcarriers are activated in the Frequency Domain {(FD)} with the index of the active subcarriers implicitly conveying extra information ~\cite{ofdmim1}. In \cite{gofdmim}, Fan \textit{et al.} generalized the OFDM-IM principle for increasing the spectral efficiency. Additionally, both the Uplink (UL) and Downlink (DL) of MU OFDM-IM have been investigated in \cite{muofdmim}, where the Peak-to-Average Power Ratio (PAPR) has been reduced compared to that of classical MU OFDM systems.  However, subcarrier-index modulated OFDM suffers from a potential throughput reduction compared to the classic {OFDM} due to the deactivation of a number of subcarriers. Hence, Zhang \textit{et al.}~\cite{csofdmim} proposed an improved {SIM} concept relying on Compressed Sensing ({CS})~\cite{cs}, which benefits from the sparsity of symbols in the {FD} by compressing the sparse transmit vector.\par
Recently, the concept of IM has also been applied to multiple dimensions, which we refer to as Multi-dimensional IM (MIM)~\cite{sfim,mflsm,mim1,mim2,chaomim}. CS techniques can also be harnessed in the MIM scheme. For instance, in \cite{csmim} a CS-aided MIM scheme was conceived, which applied IM both in the frequency and spatial-domain. Then, in\cite{jmim}, a CS-aided joint MIM system has been proposed, which jointly designed multiple dimensions and harnessed the CS technique. Furthermore, applying the MIM in Large Scale MU-MIMO (LS-MU-MIMO) scenarios is promising. Hence,CS-aided Space-Frequency Index Modulation (CS-SFIM) was proposed in \cite{mimmu} for LS-MU-MIMO-UL scenarios. \textit{However, a significant open challenge for  LS-MU-MIMO systems is the design of a reliable, yet low-complexity Multi-User Detector (MUD) for employment at the Base Station (BS).} \par
Conventional linear detectors, such as the MMSE detector, are effective in MU-OFDM scenarios, but their performance is far from that of the complex exhaustive-search-based Maximum Likelihood (ML) detector~\cite{muofdmim}. However, ML detectors may only be harnessed for systems supporting a low number of users due to their potentially excessive computational complexity. As a remedy, non-linear detectors such as Message Passing (MP) schemes~\cite{mpcs}\cite{mimoamp} achieve a good performance at a reasonable complexity for MIMO systems. MP-based detectors rely on Gaussian distributions for approximating the posterior probability distribution of the transmitted symbols conditioned on the received signal. Hence, the Approximate Message Passing (AMP) \cite{mpcs} detector has been considered both in SM \cite{ampgsmofdmim} and OFDM-IM \cite{ampofdmim} schemes. Then, Chakrapani \textit{et al.}~\cite{gsfimamp} proposed a MP-based low-complexity detection method for reducing the complexity of {SFIM} detection. If we extend the single user scenario  to massive MIMO and MU-MIMO schemes,  a high MUI is introduced, where the AMP \cite{mpcs} detector performs poorly in the presence of imperfect Channel State Information (CSI). These problems have  been partially resolved by the Orthogonal AMP (OAMP) detector~\cite{oamp} that integrates the AMP with the linear MMSE filtering. As a further design alternative, Expectation Propagation (EP) \cite{ep}\cite{epmimo} based detectors rely on a posterior distribution approximation based on the independent Gaussian approximation and the Gaussian cavity distribution in the detection process. Hence, they can significantly outperform the OAMP detector, as a benefit of harnessing regularization parameters into the MMSE filter, which are iteratively adjusted based on both the CSI and on the level of MUI. However, it has been shown that the use of this posterior distribution approximation leads to the partial loss of MUI information in these detectors and a severe performance degradation in a high-MUI scenario. 
 \textit{Therefore, despite the aforementioned improvements in MP algorithms, there remains a significant performance gap compared to the ML detector.}

Neural Network (NN)-based detectors have recently been proposed in \cite{learnmimo,dlphy,dldet,gnnmimo} for striking an improved performance versus complexity trade-off in the case of imperfect CSI and MUI. To be more specific, this may be achieved by embedding the iteration process of the MP detectors into NN layers and optimizing the NN parameters during the training  phase. For example, the OAMPNet detector of \cite{learnmimo} combines the OAMP and NN to deal with potential ill-conditioned channel matrices, which results in a significant performance improvement compared to the conventional OAMP detector. Additionally, the Graph Neural Network (GNN) aided detector \cite{gnnmimo} \cite{gnnmu}  is based on the belief propagation algorithm and utilizes a pair-wise Markov Random Field (MRF) model, which significantly reduces the number of NN parameters. However,  a significant performance gap remains between the practical state-of-the-art  detectors and the ML detector in the presence of ill-conditioned channel matrices and high MUI. Then, a AMP-based GNN model was proposed in\cite{gnnampmumimo} for MU-OFDM scenarios, which significantly reduces the detection complexity, while attaining near-ML performance. Furthermore, the EP-based GNN model (GEPNet) was proposed in \cite{gnnepmimo,gnnmumimo} for achieving near-optimal performance, while slightly increasing the computational complexity compared to the GNN-AMP.\par

 \begin{table*}[!thb]
  \centering
  \caption{Contrasting our contributions to the literature }
  \scalebox{1.2}{
  \begin{tabular}{|l||c||c|c|c|c|c|c|c|c|c|c|c|} 
  \hline
  Contribution&proposed*&\cite{immu}&\cite{musm}&\cite{epsoftmu}&\cite{gnnmu}&\cite{gnnmumimo}&\cite{gnnampmumimo}&\cite{gnnepmimo}&\cite{mimmu}&\cite{muofdmim}&\cite{ismofdm}\\
  \hline
  \hline
       MU scenario&\checkmark&\checkmark &\checkmark &\checkmark &\checkmark  &\checkmark  &\checkmark &\checkmark &\checkmark&\checkmark &\checkmark \\
  \hline
  Large-scale MIMO&\checkmark&\checkmark &\checkmark & & & &&  &\checkmark &\checkmark&\checkmark\\
  \hline
  CS at the transmitter&\checkmark& & & & & & & &\checkmark &  &\\
  \hline
  
  Separate MIM&\checkmark& & & & & & & &\checkmark& &\checkmark \\
  \hline
  MMSE-based GNN detector&\checkmark& & && &\checkmark & & & &&\\
   \hline
     AMP-GNN detector&\checkmark& & && & &\checkmark & & &&\\
   \hline
     GEPnet detector&\checkmark& & && & & &\checkmark & &&\\
   \hline

  \end{tabular}
  }
  \label{Table:ref}
\end{table*}

Against the above backdrop, in this paper, we propose a CS-SFIM scheme for the LS-MU-MIMO-UL, which combines the benefits of LS-MU-MIMO, SM, GSM, OFDM-IM and CS for UL transmission over frequency-selective fading channels. In our proposed solution, we design the index of the TAs and the subcarriers separately for attaining higher implementation flexibility than the joint space-frequency indexing of \cite{sfim}, while the Space-Frequency (SF) domain indexing is considered as a joint SF matrix for factor graph design.  Then, we design an MP-based reduced-complexity MUD for the proposed system, which incorporates AMP and EP philosophies in order to strike an attractive Bit Error Rate (BER) versus complexity trade-off. Guided by this analysis, we propose a GNN based framework for improving the detection performance by fine-tuning the cavity distribution\footnote{In EP based detection, EP approximate the distribution with a tractable distribution, to achieve more accurate estimation compared with Gaussian distribution. Then, to further increase the accuracy of the replaced distribution, a cavity distribution is introduced to update the approximate factor of aimed distribution.}. Compared to the conventional Gaussian cavity function parameterized only by its mean and variance, the GNN based framework introduces additional parameters into the cavity function to capture more accurate MUI information~\cite{gnnmimo}. The choice of the AMP-based GNN detector is motivated by its low computational complexity, in contrast to the EP detector since it does not have to perform any matrix inversion operation.  We refer to the newly developed GEPNet detector, which can iteratively achieve an improved detection performance compared to the conventional EP \cite{ep}\cite{epmimo}.  We design both GEPNet and GNN-AMP detectors exhibiting robustness to the fluctuating number of users, which allows the proposed detectors to be trained only once even for systems supporting a time-variant number of users. Our simulation results show that the GEPNet and GNN-AMP detectors are capable of significantly outperforming the EP~\cite{ep}, the AMP~\cite{oamp} and the GNN~\cite{gnnmu}  based detectors.

\begin{figure*}[!t]
\centering
{\includegraphics[width=15.5cm]{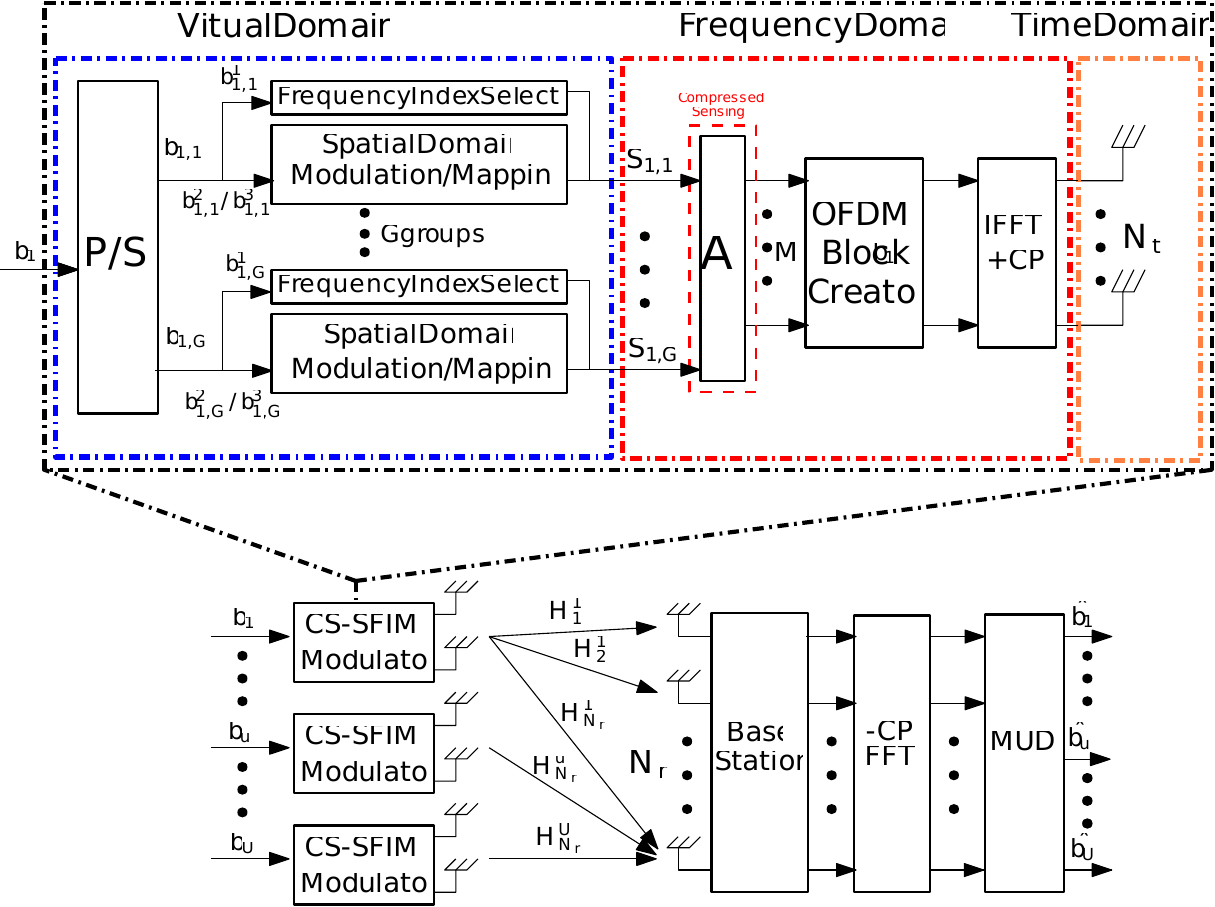}%
\caption{MU CS-SFIM UL transceiver architecture, where the BS has $N_r$ RAs to serve the $U$ UL users and each user has $N_t$ TAs.}
\label{fig:sturcture}}
\end{figure*}

In summary, Table~\ref{Table:ref} boldly contrasts the novelty of this paper to the literature. More explicitly, the contributions of this paper are elaborated on as follows:
\begin{enumerate}

    \item We propose a CS-SFIM for the LS-MIMO-MU-UL scenario to strike a compelling complexity versus performance trade-off.  For each user, the information is conveyed over the FD-IM, the Spatial-Domain (SD) IM and the classic Amplitude Phase Modulation (APM) symbols. Additionally, to attain an improved diversity gain, in contrast to the MIM scheme of \cite{csmim}, different TA activation patterns are proposed for the different transmitted APM symbols instead of constraining the active TAs for the entire subcarrier group of OFDM symbol. The received signal model can be formulated based on a SF domain matrix, which is a sparse matrix that can facilitate CS-SFIM for LS-MIMO-MU-UL.
    \item  We design MP detectors for the CS-SFIM-MU system, where we evaluate the accuracy of their joint posterior distribution approximations and analyze the relationship between this accuracy and the cavity distributions used for calculating the symbol estimates. Then, we compare the computational complexity of the MP-based detector to that of the ML and  MMSE detectors. We demonstrate that the AMP and EP detectors are capable of striking a better BER performance versus complexity trade-off than the conventional MU-MIMO-OFDM system using the ML and MMSE-based detector.
    \item As a further advance, we develop a GNN-based framework to improve the posterior distribution approximation by fine-tuning the cavity distributions in the MP detectors. The proposed GNN-based framework is designed to allow processing of the \textit{a priori} knowledge generated by the MP detectors. We propose a pair of novel GNN based detectors that use the GNN-based framework developed to improve the classic EP and AMP. The first detector, namely the AMP-GNN scheme is capable of achieving a reasonable detection performance at low complexity, while the GEPNet detector is proposed for attaining an improved detection performance by introducing iterations.
    \item We demonstrate that training the proposed GNN-based detector only once is sufficient, even if we have a variable number of users supported by the proposed system. This conclusion is verified by mathematical proof that the proposed detectors share GNN's permutation equivariance property. Our simulation results further demonstrate that the proposed GNNs are robust to changes in the number of users. By contrast, conventional Deep Neural Network (DNN) based detectors require individual training for each specific number of users. 
\end{enumerate}

The rest of the paper is organized as follows. In Section~\ref{sec:system}, we present the proposed system model. In Section~\ref{sec:det}, we present our linear and non-linear MU detectors. Then, in Section~\ref{sec:gnn} we develop our proposed GNN-based MP detector for the CS-SFIM system considered. In Section~\ref{sec:perf}, we present our simulation results, while our conclusions are offered in Section~\ref{sec:conc}.

\section{SYSTEM MODEL}
\label{sec:system}
In this section, we introduce the transceiver model of the CS-SFIM aided LS-MIMO-MU-UL supporting $U$ UL users, where each user is equipped with $N_t$ TAs communicating with a BS having $N_r$ Receive Antennas (RA), where $N_r$ is on the order of tens to hundreds. The modulation of the proposed CS-SFIM scheme at the transmitter is illustrated in Fig.~\ref{fig:sturcture}, where both Generalized OFDM-IM (GFIM) and SM techniques are harnessed for each user to improve the UL throughput and diversity gain. The specific procedures of both subcarrier and TA index modulation designs in CS-SFIM systems are detailed in the following subsections.

\begin{figure}[!htb]
  \centering
  \includegraphics[width=8cm]{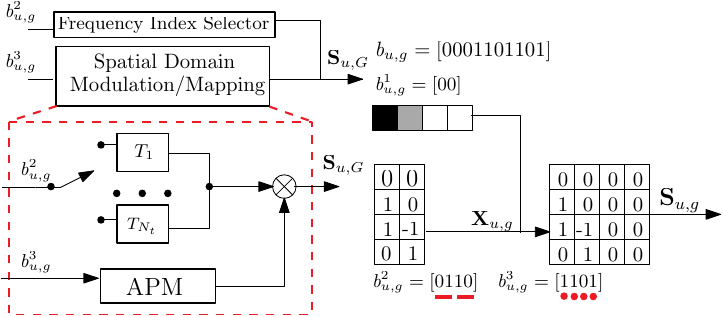}
  \caption{structure and Example of CS-SFIM modulation at $g$-th group of $u$-user with $k=2$ and bit sequence of $[0001101101]$.}
  \label{fig:sfimmod}
\end{figure} 

\subsection{UL Transmission}
As shown in Fig.~\ref{fig:sturcture}, in the block diagram of the LS-MU CS-SFIM system, there are $U$ UL users communicating with a BS, where IM is employed by each user. As shown in Fig.~\ref{fig:sturcture}, the bit sequence $\boldsymbol{b}_u$ of the $u$-th UL user is split into $G$ subcarrier groups for the CS-SFIM schemes, where all the $G$ subcarrier groups have the same modulation process. For the CS-SFIM scheme, the information sequence $\boldsymbol{b}_{u,g}$ is split into $\boldsymbol{b}_{u,g}^1$ and $\boldsymbol{b}_{u,g}^2$ and $\boldsymbol{b}_{u,g}^3$. Explicitly, $\boldsymbol{b}^3_{u,g}$ represents the APM constellation, $\boldsymbol{b}_{u,g}^1$ is conveyed by the Virtual-Domain (VD) Frequency Index Selector block and $\boldsymbol{b}_{u,g}^2$ is conveyed by the SM modulation. 
\subsubsection{Virtual-Domain Frequency Index Selection}
\label{sec:fd}
With the aid of CS techniques, we can readily extend the set of available subcarriers by introducing the VD \footnote{{VD} is the domain where the FIM is deployed. This concept was first introduced by Zhang \textit{et al.} in \cite{ZhangCompressedOFDM} in OFDM-IM systems for improving the spectral efficiency.}, where the bit sequence $\boldsymbol{b}_{u,g}^1$ of length $ \lfloor\log_2(C^K_{N_{v}})\rfloor$ is entered into the Frequency index selector of Fig.~\ref{fig:sturcture}. For each virtual subcarrier block, we consider $N_v$ available subcarriers for the GFIM modulation.
Specifically, the bit sequence $\boldsymbol{b}_{u,g}^1$  controls the subcarrier selection. A simple example is presented in Table~\ref{Table:sub} for $K=2$ and $N_v=4$. If $\boldsymbol{b}_{u,g}^1=[0\ 0]$, the first and the second subcarriers are activated for carrying information, while the other subcarriers remain inactive.

\subsubsection{Spatial-Domain Modulation/Mapping}
As shown in Fig.~\ref{fig:sturcture}, for each subcarrier group, the information sequence $\textbf{b}_{u,g}^2$ is conveyed by the SD mapping block. To increase the SE in the SD, generalized SM is utilized in the system of Fig.~\ref{fig:sturcture}, where $N_a$ out of $N_t$ TAs associated with $N_a\in(1,2,...,N_t)$ are activated by the antenna selector. The information sequence $\textbf{b}_{u,g}^2$ in the $g$-th subcarrier group of the $u$-th user has a length of $K \lfloor\log_2C(N_t,N_a) \rfloor$, where $ \lfloor\log_2C(N_t,N_a) \rfloor$ bits are needed for selecting the activated $N_a$ TAs, which are applied to $K$ activated subcarriers.
\subsubsection{Space-Frequency Index Modulation}
As shown in Fig.~\ref{fig:sturcture}, the active subcarrier of the VD frequency index selector and the active antenna of the SD antenna selector can form a SF matrix $\mathcal{S}_{u,g}\in\mathbb{C}^{N_t\times N_v}$.  More specifically, the SF matrix is obtained by assigning the column index to the SD active antenna and the row index to the active VD subcarriers, which results in $N_a\cdot K$ non-zero elements in the sparse SF matrix $\mathcal{S}_{u,g}$.

\begin{table}[!htb]
\caption{A Look-up table of subcarrier selection for CS-SFIM systems having $K=2,N_{v}=4$.}
  \centering
  \begin{tabular}{|c|c|c|} 
  \hline
 $b_{2}$&Indices&Allocation\\
  \hline
  [0 0]& (1,2)&[$K_{1}$ $K_{2}$ 0 0]\\
  \hline
  [0 1]&(2,3)&[0 $K_{1}$ $K_{2}$ 0]\\
  \hline
  [1 0]&(3,4)&[0 0 $K_{1}$ $K_{2}$]\\
  \hline
  [1 1]&(1,4)&[$K_{1}$ 0 0 $K_{2}$]\\
  \hline
  \end{tabular}

  \label{Table:sub}
\end{table}

\subsubsection{APM}
The bit sequence $\textbf{b}_{u,g}^3$ of length $N_{a}K\log_2\mathcal{L}$ is mapped to the $N_{a}K$ positions of the non-zero elements in $\mathcal{S}_{u,g}$ to create the $(N_a\cdot K)$ $\mathcal{L}$-ary APM symbols of Fig.~\ref{fig:sfimmod}. Assuming that we have $K=2,N_a=2,N_t=4,N_v=4$ and $\mathcal{L}=2$, the information sequence $\textbf{b}_{u,g}=\textbf{b}_{u,g}^1+\textbf{b}_{u,g}^2+\textbf{b}_{u,g}^3= \lfloor\log_2(C^K_{N_{v}}) \rfloor+K \lfloor\log_{2}C^{N_t}_{N_a}) \rfloor+N_{a}K\log_2\mathcal{L}$, has a length of $(\log_{2}4+2\times\log_{2}4+ 2\times2\log_22)=10$ bits. As shown in the example in Fig.~\ref{fig:sfimmod}, we have an information bit sequence of
$\textbf{b}_{u,g}=[0001101101]$.Then $\textbf{b}_{u,g}^1=[00]$ represents that the first and second subcarriers in a subcarrier group are activated as shown in Table~\ref{Table:sub}, while $\textbf{b}_{u,g}^2=[0110]$ indicates that the second and third TAs are activated for the first subcarrier and that the third and fourth antennas are activated for the second subcarrier, while the total number of activated subcarriers in each VD subcarrier group is $K=2$. Then a SF matrix $\mathcal{S}_{u,g}\in\mathbb{C}^{4\times 4}$ is constructed with the aid of $\textbf{b}_{u,g}^1$ and $\textbf{b}_{u,g}^2$, which has four non-zero elements, as shown in Fig.~\ref{fig:sfimmod}.  Lastly, $\textbf{b}_{u,g}^3=[1101]$  corresponds to allocating BPSK symbols of $(1)$,$(1)$,$(-1)$ and $(1)$ to the corresponding position of non-zero elements in the SF matrix. 


\subsubsection{Compressed Sensing and Block Assembly}
For $G$ subcarrier groups per OFDM symbol, each subcarrier group symbol is modulated by the FIM-SM block in parallel, $G$ SF matrices $\mathcal{S}_{u,g} (g=1,2,...,G)$  are established for the $u$-th user. Then, for each subcarrier group of the $\tau$-th TA of the $u$-th user,  the CS measurement matrix $\textbf{A}\in\mathbb{C}^{N_f\times N_v}$ is harnessed for compressing the $N_v$-dimensional vector $\mathcal{S}_{u,g}^\tau(\tau=1,2,...,N_t)$ gleaned in the VD into a $N_f$-dimensional vector $\textbf{s}_{u,g}^\tau$ from the $\tau$-th TA in the FD, which can be expressed as:

\begin{equation}
\textbf{s}_{u,g}^{\tau}=\textbf{A}\mathcal{S}_{u,g}^{\tau}.
\label{equation:cs}
\end{equation}
Then, the FD signals $\textbf{s}_{u,g}$ transmitted from the $N_t$ TAs associated with $g$-th subcarrier block of the $u$-th user, where $\textbf{s}_{u,g}\in\mathbb{C}^{N_t\times N_f}$. Afterwards, the Block Assembler of Fig.~\ref{fig:sturcture} gathers $G$ FD signals $\textbf{s}_{u,g}$ to form a long SF frame $\textbf{s}_{u}\in\mathbb{C}^{N_t\times N_{f}G}$, which contains $N_t$ OFDM symbols for transmission from $N_t$ TAs and can be formulated as $\textbf{s}_{u}=[\textbf{s}_{u,1},\textbf{s}_{u,2},\cdots,\textbf{s}_{u,G}]$. Additionally, the entire OFDM has $N_c$ subcarrier number, which meets the condition that $N_c=N_f\times G$. Then, Inverse Fast Fourier transform (IFFT) is applied to the OFDM symbol at each TA to obtain the Time-Domain (TD) symbol, which is then extended by the Cyclic Prefix (CP). Afterwards, the TD signals of the $u$-th user are transmitted from $N_t$ TAs over the wireless channel. Hence, for the LS-MU-MIMO-UL scenario, the maximum achievable rate of the proposed CS-SFIM system in bits/s/Hz can be expressed as \cite{mimmu}:
\begin{equation}
R_t=U \times \frac{G(\log_2(C^K_{N_{v}})+K\log_{2}C^{N_t}_{N_a}+]+N_{a}K\log_2\mathcal{L})}{N_c+N_{cp}}.
\label{equation:rate}
\end{equation}

To achieve the theoretical maximum transmission rate, a powerful detector is required in this scenario at an affordable complexity.

\section{Multi-User Detection}
\label{sec:det}

In Fig.~\ref{fig:sturcture}, the MU-UL signals are superimposed at the BS and received through distinct MIMO channels. Despite simultaneous transmission, user separability is achieved due to the unique SF signatures imposed by each user's channel\cite{mumimoofdm}\cite{munonlinear}. The MIMO channel model of $U$ UL users can be expressed as $\textbf{H}^u\in\mathbb{C}^{N_r \times N_t}(u=1,2,\cdots,U))$, where $\textbf{H}^u_i[n]\in\mathbb{C}^{1\times N_t}(i=1,2,\cdots,N_r, u=1,2,\cdots,U, n= 1,2,\cdots,N_c)$ denotes the complex channel gain of the $n$-th subcarrier received from the $N_t$ TAs of the $u$-th user at the $i$-th RA. Assuming perfect synchronization\footnote{While the synchronization among multiple users (MUs) is an important practical consideration, it is beyond the scope of this paper. Several techniques addressing MU synchronization have been proposed in the literature, including \cite{musyn1}\cite{musyn2}.}, then the FD  signal $\textbf{y}^u_i[n]$ received from the $u$-th user at the $i$-th RA after CP removal and Fast Fourier Transform (FFT)\footnote{All users simultaneously transmit their signal in the UL from $N_t$ TAs, which impinge on multiple RAs at the BS. The cyclic prefix (CP) removal and FFT based demodulation operations are then applied jointly to these stacked signals, rather than to each user’s signal individually.} is given by 

\begin{equation}
\textbf{y}^u_i[n]=\textbf{H}^u_i[n]\textbf{s}_u[n]+\textbf{w}_i^u[n],
\label{equation:rec1}
\end{equation}

where $ \textbf{s}_u\in\mathbb{C}^{N_t\times 1}$ denotes the FD symbols transmitted on the $n$-th subcarrier of the $N_t$ TAs and $\textbf{w}_i^u[n]$ is the Additive White Gaussian Noise (AWGN) with zero mean and variance of $\sigma^2$.
Then we can express the signal $\textbf{y}_i^u \in \mathbb{C}^{N_c \times 1}$ received by the $i$-th RA from the $u$-th user as:
\begin{equation}
\textbf{y}^u_i=\textbf{H}^u_i\textbf{s}_u+\textbf{w}_i^u,
\label{equation:rec2}
\end{equation}
where $\textbf{H}^u_i=\text{diag}\{\textbf{H}^u_i[n]\},(n=1,2,\cdots,N_c)$ $\textbf{H}^u_i \in \mathbb{C}^{N_c \times (N_t N_c)}$, denotes the diagonal structure of the complex channel matrix from the $N_t$ TA to the $i$-th RA. Furthermore, $\textbf{s}_u\in\mathbb{C}^{(N_tN_c) \times 1}=[(\textbf{S}_u[1])^T(\textbf{S}_u[2])^T\cdots (\textbf{S}_u[N_c])^T]^T$ represents the FD symbols vector of length $N_c$ for the $u$-th user transmitted from the $N_t$ TAs, which is sparse with $(N_a \cdot K \cdot G)$ non-zero elements. Moreover, $\textbf{w}_i^u\in\mathbb{C}^{N_t \times 1}$ denotes the AWGN vector, with $\mathcal{CN}(0,\sigma_n^2)$.
Then, the FD symbols $\textbf{y}_i\in\mathbb{C}^{N_c\times 1}$ received at the $i$-th RA 
are expressed as:
\begin{equation}
\textbf{y}_i=\sum^U_{u=1}\textbf{y}_i^u=\textbf{H}_i\textbf{s}+\textbf{w}_i,
\label{equation:eachuserdet}
\end{equation}
where we have $\textbf{H}_i=[\textbf{H}_i^1, \textbf{H}_i^2 ,\cdots ,\textbf{H}_i^U]$, $\textbf{s}=[(\textbf{s}_1)^T, (\textbf{s}_2)^T, \cdots ,(\textbf{s}_U)^T]\in\mathbb{C}^{(N_tN_cU)\times 1}$, and $\textbf{w}_i=\sum_{u=1}^U w_i^u$.
For the overall MIMO model, the signal $\textbf{y}\in\mathbb{C}^{(N_rN_c)\times 1}$ received over $N_r$ RAs at the BS can be modeled by
\begin{equation}
\textbf{y}=\textbf{H}\textbf{s}+\textbf{w},
\label{equation:alluserdetsig}
\end{equation}
where we have $\textbf{y}=[\textbf{y}_1^T \textbf{y}_2^T \cdots \textbf{y}_{N_{r}}^T]^T$, $\textbf{H}=[\textbf{H}_1^T \textbf{H}_2^T \cdots \textbf{H}_{N_r}^T]^T\in\mathbb{C}^{(N_rN_c)\times (N_tN_cU)} $ and $\textbf{w}=[\textbf{w}_1^T \textbf{w}_2^T \cdots \textbf{w}_{N_{r}}^T]^T \in\mathbb{C}^{(N_rN_c)\times 1}$.
As shown in Fig.~\ref{fig:sturcture}, the FD signal $y$ can be split into $G$ groups. Hence, the sub-group signal $\textbf{y}_g\in\mathbb{C}^{(N_rN_f)\times 1}$ of the $g$-th subcarrier group of the $U$ users received from the $N_t$ TAs over the $N_r$ RAs at the BS can be expressed as: 
\begin{equation}
\textbf{y}_g=\textbf{H}_g\textbf{s}_g+\textbf{w}_g.
\label{equation:alluserdetsig2}
\end{equation}
In (\ref{equation:alluserdetsig2}) the $g$-th subcarrier group $\textbf{s}_g\in\mathbb{C}^{(N_tN_fU)\times 1}$ denotes the $g$-th subcarrier group FD signals transmitted from the $N_t$ TAs of the $U$ users. Additionally, $\textbf{w}_g\in\mathbb{C}^{(N_rN_fU)\times 1}$ is the Gaussian noise vector. 
The equivalent channel matrix $\textbf{H}_g\in\mathbb{C}^{(N_rN_f)\times (N_tN_fU)}$ corresponding to the $g$-th subcarrier group of the $U$ users transmitted signals from the $N_t$ TAs over the $N_r$ RAs can be expressed as:

\begin{equation}
\textbf{H}_g = \begin{bmatrix}
\textbf{H}_{1,g}^1 &\textbf{H}_{1,g}^2&\dots& \textbf{H}_{1,g}^{U} \\
\textbf{H}_{2,g}^1 &\textbf{H}_{2,g}^2&\dots& \textbf{H}_{2,g}^{U} \\
\vdots&\vdots&\ddots&\vdots\\
\textbf{H}_{N_r,g}^1 &\textbf{H}_{N_r,g}^2&\dots& \textbf{H}_{N_r,g}^{U}
\end{bmatrix},
\end{equation}
where $\textbf{H}_{i,g}^u\in\mathbb{C}^{N_f\times (N_tN_f)}$, represents the complex-valued channel from $N_t$ TAs to the $i$-th RA of the $u$-th user received. Then, $\textbf{H}_{i,g}$ can be written as:
\begin{equation}
\textbf{H}_{i,g}^u = \begin{bmatrix}
\textbf{H}_{i,g}^u[1] &0&\dots& 0 \\
0&\textbf{H}_{i,g}^u[2]&\dots&0 \\
\vdots&\vdots&\ddots&\vdots\\
0 &0&\dots&\textbf{H}_{i,g}^u[N_f]
\end{bmatrix},
\end{equation}
where $\textbf{H}_{i,g}^u[n] \in\mathbb{C}^{1\times N_t},n=1,2,\cdots,N_f$ represents the channel of the $n$-th FD subcarrier in the $g$-th subcarrier group of the $u$-th user transmitted from the $N_t$ TAs to the $i$-th RA.
Then, the signals $\textbf{s}_g$ in (\ref{equation:alluserdetsig2}) transmitted to the $g$-th group can be expressed as:
\begin{equation}
\textbf{s}_g=\bar{\textbf{A}}\textbf{I}_{SI}^g\textbf{I}_{AC}^g\textbf{X}_L^g+\textbf{w}_g,
\label{equation:expandedsignal1}
\end{equation}
where $\bar{\textbf{A}}\in\mathbb{C}^{(N_tN_fU)\times (N_tN_vU)}$ denotes the equivalent\footnote{In this case, $\bar{\textbf{A}}$ is  a larger matrix than $\textbf{A}$, which is a grouped measurement matrix for compressing $U$ users' signal transmitted from $N_t$ TAs.} CS measurement matrix of $\textbf{A}$ in ($\ref{equation:cs}$), $\textbf{I}_{SI}^g\in\mathbb{C}^{(N_tN_vU)\times (N_tKU)}$ represents the subcarrier index activation pattern, while $\textbf{I}_{AC}^g\in\mathbb{C}^{(N_tKU)\times (KN_aU)}$ represents the equivalent TA index activation pattern for $U$ users in the $g$-th subcarrier group.\par
Furthermore, $\textbf{X}_{L}^g\in\mathbb{C}^{(KN_aU)\times 1}$ in (\ref{equation:expandedsignal1}) represents the vector, which has $KN_aU$ APM symbols in the $g$-th subcarrier group for $U$ users. Therefore, we can extend (\ref{equation:alluserdetsig2}) as:

\begin{equation}
\textbf{y}_g=\textbf{H}_g\bar{\textbf{A}}\textbf{I}_{SI}^g\textbf{I}_{AC}^g\textbf{x}_L^g+\textbf{w}_g=\textbf{H}_g\bar{\textbf{A}}\textbf{I}_g\textbf{X}_L^g+\textbf{w}_g,
\label{equation:finalextend}
\end{equation}
where $\textbf{I}_g\in\mathbb{C}^{(N_tN_vU)\times (KN_aU)}=\textbf{I}_{SI}^g\textbf{I}_{AC}^g$ is the equivalent SF matrix containing the information of the active TAs and the active subcarriers of $U$ users in the $g$-th subcarrier group.

Then, we consider the case that not all users are active at a time. We introduce a activity indicator $\rho_u\in{1,0}$ and modify (\ref{equation:eachuserdet}) as
\begin{equation}
\textbf{y}_i=\sum^U_{u=1}\rho_u\textbf{y}_i^u=\textbf{H}_i\textbf{s}+\textbf{w}_i,
\label{equation:random1}
\end{equation}

and (\ref{equation:alluserdetsig}) can be modified as 
\begin{equation}
\textbf{y}=\textbf{H}\boldsymbol{I_{\rho}}\textbf{s}+\textbf{w},
\label{equation:random2}
\end{equation}
\noindent and for each subcarrier group
\begin{equation}
\textbf{y}_g=\textbf{H}_g\boldsymbol{I_{\rho}}\bar{\textbf{A}}\textbf{I}_g\textbf{X}_L^g+\textbf{w}_g,
\label{equation:random3}
\end{equation}
where $\boldsymbol{I_{\rho}}$ indicates the active users, which is decided by $K_a$.\\

In the following, we will detail both conventional detectors and learning-based detectors.

\subsection{Conventional Detectors}
\label{sec:conv}

Conventional exhaustive-search based ML detection can be applied at the receiver, albeit this may lead to excessive complexity~\cite{imconcept}. In the following section, we present the conventional {ML}-based Hard-Decision ({HD}) detector, followed by the MMSE detector. Then, the MP-based AMP and EP detectors are discussed, followed by the proposed GNN-based methods, where the GNN is applied to improve the MP and AMP detector performance iteratively.

\subsubsection{Maximum Likelihood Detection}
As shown in Fig.~\ref{fig:sturcture} and  Fig.~\ref{fig:sturcture}, the information sequence $\textbf{b}_u$ $(u=1,2,\cdots,U)$ is processed group by group by the CS-SFIM modulator. Then the signal received at the BS is detected by the MU detector of Fig.~\ref{fig:sturcture}.  To simplify the analysis, we only introduce the ML-aided detector for the $g$-th subcarrier group, which can be generalized to the entire received signal.
Then, we assume that $\langle\alpha_g,\beta_g,\gamma_g\rangle$ represents the information in the $g$-th subcarrier group for $U$ users of  the CS-SFIM system shown in Fig.~\ref{fig:sturcture}. Here, $(\textbf{X}_L^g)_{\alpha_{g}}$ denotes the $\alpha_g$-th realization of the $KN_aU$ APM symbols for $\alpha_g=1,2,\cdots,L^{KN_aU}$. Furthermore, $(\textbf{I}_{SI}^g)_{\beta_g}$ represents the $\beta_g$-th indexing pattern of the active subcarrier of the $U$ users, where $\beta_g=1,2,\cdots,N^{U}_{SI}$. $(\textbf{I}_{AC}^g)_{\gamma_g}$ denotes the $\gamma_g$-th active TA selection pattern for the $KU$ active subcarriers in the $g$-th subcarrier group of $U$ users, where we have $\gamma_g=1,2,\cdots,(N_{AC}^K)^U$. Correspondingly, we use $\langle\hat{\alpha}_g,\hat{\beta}_g,\hat{\gamma}_g\rangle$ to represent the estimation of the input information. Hence, we have the optimal HD ML detector based on (\ref{equation:alluserdetsig2}) and (\ref{equation:finalextend}) as:

\begin{equation}
\begin{split}
\langle\hat{\alpha},\hat{\beta},\hat{\gamma}\rangle&=\arg\min_{\alpha,\beta,\gamma} \| \textbf{Y}_g-\textbf{H}_g\bar{\textbf{{A}}}
(\textbf{I}^g_{SI})_\alpha)(\textbf{I}^g_{AC})_\beta(\textbf{X}^g_L))_\gamma \|^2,\\
&=\arg\min_{\alpha,\beta,\gamma} \| \textbf{Y}_g-\textbf{H}_g\bar{\textbf{{A}}}
 \hat{\textbf{{s}}}\|^2,\\
\end{split}
\end{equation}
where $\hat{\textbf{s}}=(\textbf{I}^g_{SI})_\alpha)(\textbf{I}^g_{AC})_\beta(\textbf{X}^g_L))_\gamma$ denotes the search space of the $g$-th subcarrier group. The ML detector achieves the optimal detection performance with excessive complexity that increases exponentially with throughput.

\subsubsection{MMSE}
Based on the signal $\textbf{y}_g$ of $g$-th subcarrier group in (\ref{equation:alluserdet2}) and (\ref{equation:finalextend}) received from all users, the signal of the $g$-th subcarrier group can be reformulated as:
\begin{equation}
\textbf{y}_g=\bar{\textbf{H}}_g\textbf{x}_g+\textbf{w}_g,
\label{equation:alluserdet2}
\end{equation}
where $\bar{\textbf{H}}_g\in\mathbb{C}^{(N_rN_f)\times (N_tN_vU)}=\textbf{H}_g\bar{\textbf{A}}$ is the equivalent channel matrix of all users in the $g$-th subcarrier group and $\textbf{x}_g\in\mathbb{C}^{N_tN_vU\times 1}$ is the equivalent information signal transmitted by all users over the $g$-th subcarrier group and modulated by TA index, subcarrier index and APM.

Then, the Linear MMSE (LMMSE) detector is applied to the received signal $\textbf{y}_g$ of ($\ref{equation:alluserdet2}$) for mitigating the MU interference and acquiring the estimator $\hat{\textbf{x}}_g\in\mathbb{C}^{(N_tN_vU)\times 1}$ of the transmitted signal $\textbf{x}_g$ of ($\ref{equation:alluserdet2}$), which can be expressed as: 
\begin{equation}
\hat{\textbf{x}}_g=(\bar{\textbf{H}}_g^H\bar{\textbf{H}}_g+\frac{1}{\lambda}\textbf{I}_{(N_tN_fU)})^{-1}\bar{\textbf{H}}_g^H\textbf{y}_g,
\label{eq:mmsedet}
\end{equation}
where $\lambda=\textit{E}\|\bar{\textbf{H}}_g^H\textbf{x}\|^2_2/\textit{E}\|\textbf{w}_g\|^2_2$ represents the average SNR per symbol and is expressed as $\lambda=\frac{1}{\sigma^2}$. Afterwards, the estimated signal $\hat{\textbf{s}}_g$ is split into $U$ vectors in parallel, and transformed into matrix form as $[\hat{\textbf{x}}_g^1~\hat{\textbf{x}}_g^2~\cdots ~\hat{\textbf{x}}_g^U]^T \in \mathbb{C}^{U\times N_tN_f}$, where $\hat{\textbf{x}}_g^u\in\mathbb{C}^{(N_tN_f)\times 1}(u=1,2,\cdots,U)$ is the MMSE estimate of the $u$-th users' signal. Based on (\ref{equation:finalextend}) and (\ref{eq:mmsedet}), the signal $\hat{\textbf{x}}_g^u$ estimated for the $u$-th user can be expressed as:
\begin{equation}
\hat{\textbf{x}}_g^u=\textbf{I}_{g}^u\textbf{X}_L^{g,u}+\hat{\textbf{w}}_g^u,
\label{equation:expandedsignal3}
\end{equation}
where $\textbf{I}^u_g\in\mathbb{C}^{N_tN_v\times KN_a}=\textbf{I}_{SI}^{g,u}\textbf{I}_{AC}^{g,u}$ denotes the joint equivalent index pattern of the SF matrix. It contains the activation patterns  $\textbf{I}_{AC}^{g,u}\in\mathbb{C}^{N_tN_v\times N_tK}$ of TAs and subcarriers $\textbf{I}_{SI}^{g,u}\in\mathbb{C}^{N_tK\times KN_a}$ arriving from the $u$-th user in the $g$-th subcarrier group. Furthermore, $\textbf{X}_{L}^{g,u}\in\mathbb{C}^{KN_a\times 1}$ represents the APM symbol vector of the $u$-th user and $\hat{\textbf{w}}_{g}^{u}\in\mathbb{C}^{N_tN_v\times 1}$ is the AWGN vector.
Finally, we can demodulate $\hat{\textbf{x}}_g^u$ without encountering MUI, which is suppressed in (\ref{eq:mmsedet}).

\subsection{Maximum a Posteriori Detector and Factor Graph}
Given the received signal $\textbf{y}$ and fading matrix $\bar{\textbf{H}}$ of (\ref{equation:alluserdet2}), the \textit{a posteriori} probability $p(\textbf{x}|\textbf{y})$ is used for demodulating \textbf{x} based on the received signal \textbf{y} as:
\begin{equation}
\hat{\textbf{x}}=\arg\max_{\textbf{x}\in\mathcal{S}^{\Omega}} p(\textbf{x}|\textbf{y}),
\label{eq:posteriori}
\end{equation}
where $\Omega=N_tN_vU$ is the size of the transmit signal vector $\textbf{x}$ of a specific subcarrier group. To simplify the calculation, we assume that $\textbf{y}$ corresponds to a single subcarrier group.
The $i$-th element $x_i$, $1\leqslant i\leqslant \Omega$, of $\textbf{x}$ can be estimated by the \textit{a posteriori} Probability Mass Function (PMF). Given (\ref{eq:posteriori}), we have

\begin{equation}
\hat{x}_i=\arg\max \sum_{\textbf{x}\in\mathcal{S}^{\Omega}} p(\textbf{x}|\textbf{y}).
\label{eq:pmf}
\end{equation}
According to Bayes' theorem, we can have 
\begin{equation}
p(\textbf{x}|\textbf{y})=\frac{p(\textbf{y}|\textbf{x})p(\textbf{x})}{p(\textbf{y})}\varpropto p(\textbf{y}|\textbf{x})p(\textbf{x}),
\label{eq:probrelation}
\end{equation}
where $p(\textbf{x})=\prod^{\Omega}_{i=1}p(x_i)$ is the joint $\textit{a priori}$ probability function of all the symbols in the SF matrix domain and $p(\textbf{y})=\sum_{\textbf{x}\in\mathcal{S}^{\Omega}}p(\textbf{y}|\textbf{x})$ is the PDF  of the received signal $\textbf{y}$ represented as:\par
\begin{equation}
p(\textbf{y}|\textbf{x})=\prod^{\Phi}_{j=1}p(y_j|\textbf{x}),
\label{eq:probyx}
\end{equation}
where $\Phi=N_rN_f$ is the size of the received signal vector $\textbf{y}$ and $p(y_j|\textbf{x}),1\leqslant j\leqslant \Phi$, is the conditional probability of the $j$-th observation $y_j$, given the transmitted symbol $\textbf{x}$. Bearing in mind the sparsity of the index distribution in the SF domain matrix, each symbol is interfered by  $(L-1)$ symbols  of $L$ active index elements in the SF matrix.
By combining  (\ref{eq:probrelation}) and (\ref{eq:probyx}) into (\ref{eq:pmf}), we obtain
\begin{equation}
\hat{x}_i=\arg\max_{x_i\in\mathcal{S}} \sum_{\textbf{x}\in\mathcal{S}^{\Omega(\Omega-1)},x_i} p(\textbf{x})\prod_{j\in\mathcal{R}_i}p(y_j|\textbf{x}),
\label{eq:pmf2}
\end{equation}
where $\mathcal{R}_i$ represents the non-zero element set for $\bar{\textbf{H}}$ and $p(y_j|\textbf{x})$ is formulated as
\begin{equation}
p(y_j|\textbf{x})=\frac{1}{\pi\sigma^2}\exp(\frac{-| y_j-\bar{\textbf{H}}\textbf{x}|^2}{\sigma^2}).
\label{eq:cdf}
\end{equation}

\begin{figure}
  \centering
  \includegraphics[width=6.5cm]{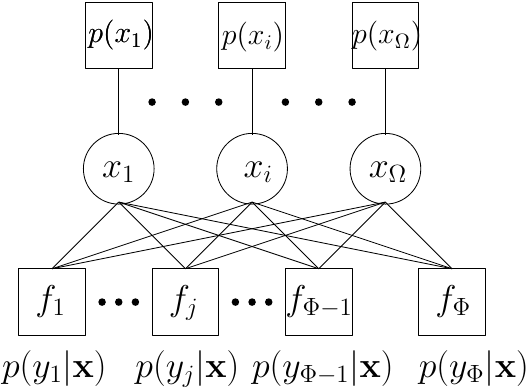}
 \caption{Factor graph for CS-SFIM}
  \label{Figure:fact}
\end{figure}

The factor graph of the CS-SFIM system is illustrated as shown in Fig.~\ref{Figure:fact},  where the SF signal $\textbf{x}$ is represented by a vector and the non-zero elements ${x}_i$ are mapped to the $i$-th Variable Node (VN) associated with the circles. Then the received signal $y_j$ is treated as the $j$-th Factor Node (FN) with solid represented by the squares in Fig.~\ref{Figure:fact}. Hence, we can describe the factorization of the joint distribution as:
\begin{equation}
p(\textbf{x},\textbf{y})=p(\textbf{y}|\textbf{x})p(\textbf{x})=\prod^{N_rN_f}_{n}f_j(y_j|\textbf{x})\prod^{\Omega}_{i}p(\textbf{x}_i),
\label{eq:jointdis}
\end{equation}
where the likelihood function $f_j(y_j|\textbf{x})$ associated with the observation vector $\textbf{y}$ is formulated as:
\begin{equation}
f_j(y_j|\textbf{x})\varpropto p(y_j|\textbf{x})\varpropto \exp(\frac{-| y_j-\bar{\textbf{H}}\textbf{x}|^2}{\sigma^2}).
\label{eq:likelihoodfunc}
\end{equation}
Based on the factor graph,  for the $i$-th element, let $\mu_{x_i\rightarrow y_j}^t(\textbf{x}_i)$ and $\mu_{y_j\rightarrow x_i}^t(\textbf{x}_i)$ denote the message sent from VN $x_i$ to FN $y_j$ and in the opposite direction at the $t$-th iteration, respectively. Then, following the principle of the sum-product algorithm \cite{sumprod}, the message passing rule can be formulated as:
\begin{equation}
\mu_{x_i\rightarrow f_j}^t(\textbf{x}_i)=\phi_{\phi_i\rightarrow x_i}(\textbf{x}_i)\prod_{j'\neq j}\mu_{f_{j'}\rightarrow x_i}^{t-1}(\textbf{x}_i),
\label{eq:forward}
\end{equation}
\begin{equation}
\mu_{f_{j}\rightarrow x_i}^t(\textbf{x}_i)=\sum_{\textbf{x}\setminus\textbf{x}_i}f_j(y_j|\textbf{x})\prod_{i'\neq i}\mu_{x_{i'}\rightarrow f_j}^{t}(\textbf{x}_i),
\label{eq:back}
\end{equation}
where $\textbf{x}\setminus\textbf{x}_i$ denotes all the enumerations of $\textbf{x}$ except for $x_i$ and $\phi_{\phi_i\rightarrow x_i}(\textbf{x}_i)=p(\textbf{x}_i)$ represents the \textit{a priori} probability of the transmitted symbol vector.\par

\subsubsection{AMP-Aided Detection}
It has been shown in~\cite{mpcs} that the {AMP} algorithm benefits from low complexity and rapid convergence. Inspired by the application of AMP in {OFDM-IM}\cite{ofdmimamp} and {GSM}\cite{smamp}, we harness the {AMP} detector over the {CS-SFIM} scheme to reduce the detection complexity compared to the full search ML and to acquire an improved performance compared to that of the MMSE detector. The AMP-based detector aims for iteratively decoupling the \textit{a posteriori} probability $p(\textbf{x}|\textbf{y})$ into a series $p(x_i|\textbf{y})$. \par

To simplify the calculations, we approximate the distribution of information  $\mu_{x_i\rightarrow y_j}^t(\textbf{x}_i)$ by a Gaussian distribution  and assume the PDF follows the  $\mathcal{N}(\textbf{x}:\textbf{r},\boldsymbol{\Sigma})$~\cite{Gaussianapp}, which can be expressed as:
\begin{equation}
\mu_{x_i\rightarrow f_j}^t(\textbf{x}_i)\varpropto p(x_i)\mathcal{N}(x_i;r^{t-1}_{{x_i\rightarrow f_j}},\Sigma^{t-1}_{x_i\rightarrow f_j}),
\label{eq:forwardgaussian}
\end{equation}
where $r^{t-1}_{{x_i\rightarrow f_j}}$ and $\Sigma^{t-1}_{x_i\rightarrow f_j}$ are the mean and variance of the Gaussian variable $x_i$.

Furthermore, the information flowing from FN to VN $\mu_{f_{j}\rightarrow x_i}^t(\textbf{x}_i)$ can also be approximated by a Gaussian PDF as:

\begin{equation}
\mu_{f_{j}\rightarrow x_i}^t(\textbf{x}_i)\varpropto p(x_i;\frac{y_j-\mathcal{Z}^{t-1}_{f_{j}\rightarrow x_i}}{h_{ji}},\frac{\sigma^2+\mathcal{V}^{t-1}_{f_{j}\rightarrow x_i}}{|h_{ji}|^2}),
\label{eq:backgaussian}
\end{equation}
where the mean is $\hat{x}^{t-1}_{{f_j\rightarrow x_i}}=\frac{y_j-\mathcal{Z}^{t-1}_{f_{j}\rightarrow x_i}}{h_{ji}}$ and the variance is $\hat{v}^{t-1}_{{f_j\rightarrow x_i}}=\frac{\sigma^2+\mathcal{V}^{t-1}_{f_{j}\rightarrow x_i}}{|h_{ji}|^2}$.
With the consideration of both forward and backward oriented information, the AMP algorithm can be further simplified as described in \cite{ampsimp}, which is presented in \textbf{Algorithm ~\ref{alg:amp}}.

\begin{algorithm}[!ht]
\DontPrintSemicolon
  
  \KwInput{$\textbf{y},\bar{\textbf{H}},\sigma^2,p(x_i)$}
  \KwInitialization{$\hat{x}^{t=1}_i=0,\hat{v}^{t=1}_i=1,Z^{t=0}_j=y_j$}
 \KwIteration{}
  \For{$t=1,2,\dots,T$}
{
    \[ \mathcal{V}^{(t)}_j=\sum|h_{ji}|^2\hat{v}_i^t
    \numberthis \label{eq:amp1}\]

    \[    \mathcal{Z}^{(t)}_j=\sum h_{ji}\hat{x}_i^t-\frac{\mathcal{V}_j^t(y_j-\mathcal{Z}_j^{(t-1)})}{\sigma^2+\mathcal{V}_j^{(t-1)}}
    \numberthis \label{eq:amp2}
    \]
    
    \[
    \hat{\Sigma}^{(t)}_i=\left( \sum \frac{|h_{ji}|^2}{\sigma^2+\mathcal{V}_j^{(t-1)}}\right)^{-1}
    \numberthis \label{eq:amp3}
    \]

    \[
    \hat{r}^{(t)}_i=\hat{x}^{(t)}_i+ \hat{\Sigma}^{(t)}_i\sum\frac{h_{ji}^*y_j-\mathcal{Z}_j^{(t)}}{\sigma^2+\mathcal{V}_j^{(t)}}
    \numberthis \label{eq:amp4}
    \]

    \[
    \hat{x}^{(t+1)}_i=\mathbb{E}(x_i;\hat{r}_i^t,\hat{\Sigma}_i^t)
    \numberthis \label{eq:ampmean}
    \]
    
    \[
    \hat{v}^{(t+1)}_i=\text{Var}(x_i;\hat{r}_i^t,\hat{\Sigma}_i^t)
    \numberthis \label{eq:ampvar}
    \]
}
\KwOutput{$\hat{\textbf{x}}^{(T)}$}

\caption{AMP Detector}
\label{alg:amp}
\end{algorithm}
Finally, (\ref{eq:ampmean}) and (\ref{eq:ampvar}) can be further expressed with the aid of the $\mathcal{L}$-APM set $\mathcal{Q}=\{q_1,q_2,\cdots,q_k,\cdots,q_L\}$, as follows:
\begin{equation}
   \hat{x}^{(t+1)}_i=\frac{\sum_{q_k\in\mathcal{Q}}q_k\mathcal{N}_{\mathbb{C}}(q_k;r_n^{(t)},\Sigma_n^{(t)})p(q_k)}{\sum_{q_k\in\mathcal{Q}}\mathcal{N}_{\mathbb{C}}(q_k;r_n^{(t)},\Sigma_n^{(t)})p(q_k)},
    \label{eq:xmeanqam}
\end{equation}
\begin{equation}
   \hat{v}^{(t+1)}_i=\frac{\sum_{q_k\in\mathcal{Q}}|q_k|^2\mathcal{N}_{\mathbb{C}}(q_k;r_n^{(t)},\Sigma_n^{(t)})p(q_k)}{\sum_{q_k\in\mathcal{Q}}\mathcal{N}_{\mathbb{C}}(q_k;r_n^{(t)},\Sigma_n^{(t)})p(q_k)}-|\hat{x}^{(t+1)}_i|^2.
    \label{eq:xvarqam}
\end{equation}

\subsubsection{EP-Aided Detection}
As a message passing-based algorithm, the EP~\cite{epmimo} detector also has two modules: \textit{observation module} and \textit{estimation module}. 

The posterior probability distribution of the transmitted signal conditioned on the received signal in  (\ref{equation:alluserdet2}),  based on  (\ref{eq:probrelation})
can be reformulated as: 
\begin{equation}
p(\textbf{x}|\textbf{y})=\frac{p(\textbf{y}|\textbf{x})p(\textbf{x})}{p(\textbf{y})}\varpropto\underbrace{\mathcal{N}(\textbf{y}:\bar{\textbf{H}}\textbf{x},\sigma^2\textbf{I})}_{p(\textbf{y}|\textbf{x})}\underbrace{\prod^{\Omega}_{i=1}p(x_i)}_{p(\textbf{x})},
\label{eq:epprop}
\end{equation}
and the \textit{a priori} probability of the transmitted symbol vector of each subcarrier group $p(\textbf{x})$ can be simplified using an unnormalized Gaussian distribution $t(x_i|V_i,\eta_i)=exp(-\frac{1}{2}V_ix_i^2+\eta_ix_i)$ to approximate the \textit{a priori} distribution $p(\textbf{x})$, where $V_i$ and $\eta_i$ represent the parameters of the Gaussian PDF.  Then, (\ref{eq:epprop}) can be rewritten as: 
\begin{equation}
\begin{split}
\hat{p}(\textbf{x}|\textbf{y})&\varpropto p(\textbf{y}|\textbf{x})\prod^{\Omega}_{i=1} t(x_i)\\
&\varpropto \mathcal{N}(\textbf{y}:\bar{\textbf{H}}\textbf{x},\sigma^2\textbf{I}) \times \prod^{\Omega}_{i=1} \exp(-\frac{1}{2}V_ix_i^2+\eta_ix_i)\\
&\varpropto \mathcal{N}(\textbf{y}:\bar{\textbf{H}}\textbf{x},\sigma^2\textbf{I}) \times  \exp(-\frac{1}{2}\textbf{x}^T\textbf{V}\textbf{x}+\boldsymbol{\eta}^T\textbf{x}),
\label{eq:epapproximate1}
\end{split}
\end{equation}
where we have $\boldsymbol{\eta}=[\eta_1,\eta_2,\cdots,\eta_{N_tN_fU}]^T$ and $\textbf{V}=\text{diag}\{[V_1,V_2,\cdots,V_{N_tN_fU}]\}$. Furthermore, by multiplying (\ref{equation:alluserdet2}) with the pseudo-inverse matrix of $\textbf{H}$ as $\bar{\textbf{H}}^{\dagger}=(\bar{\textbf{H}}^t\bar{\textbf{H}})^{-1}\bar{\textbf{H}}^T$, we arrive at $p(\textbf{y}|\textbf{x})\varpropto\mathcal{N}(\textbf{x}:\bar{\bar{\textbf{H}}}^{\dagger}\textbf{y},\sigma^2(\bar{\textbf{H}}^T\bar{\textbf{H}})^{-1})$ for a given vector $\textbf{y}$ \cite{epmimo}. To iteratively approximate $p(\textbf{y}|\textbf{x})$,  we reformulate (\ref{eq:epapproximate1}) as:
\begin{equation}
\begin{split}
\hat{p}(\textbf{x}|\textbf{y})&\varpropto \mathcal{N}(\textbf{x}:\bar{\textbf{H}}^{\dagger}\textbf{y},\sigma^2(\bar{\textbf{H}}^T\bar{\textbf{H}})^{-1}) \times \mathcal{N}(\textbf{x}:{\eta_iV_i}^{-1},V_i^{-1})\\
&\varpropto \mathcal{N}(\textbf{x}:\boldsymbol{\mu_{ep}},\boldsymbol{\Sigma_{ep}}),
\label{eq:epapproximate2}
\end{split}
\end{equation}
where $\boldsymbol{\mu_{ep}}$ and $\boldsymbol{\Sigma_{ep}}$ denote the mean vector and covariance matrix formulated as 
\begin{equation}
\boldsymbol{\Sigma_{ep}}=\left(\sigma^{-2}\bar{\textbf{H}}^T\bar{\textbf{H}}+\textbf{V}\right)^{-1},
\label{eq:mean}
\end{equation}
and
\begin{equation}
\boldsymbol{\mu_{ep}}=\boldsymbol{\Sigma_{ep}}(\sigma^{-2}\bar{\textbf{H}}^T\textbf{y}+\boldsymbol{\eta}).
\label{eq:variance}
\end{equation}
To reduce the computational complexity,  we can have a tractable distribution $q(\textbf{x})$ to approximate the intractable distribution $p(\textbf{x}|\textbf{y})$ by minimizing the Kullback-Leibler (KL) divergence, which is equivalent to matching the moments of mean and variance for these two probability distributions.  EP then approximates $p(\textbf{x}|\textbf{y})$ by a marginal distribution $q(\textbf{x})$ and iteratively updates the mean and variance as:
\begin{equation}
q^{(t)}(\textbf{x})\varpropto\prod^{\Omega}_{i=1}\underbrace{\mathcal{N}(x_i:\mu_i^{(t)},\Sigma^{(t)}_{ii})}_{q^{(t)}(x)},
\label{eq:epapproximatedist}
\end{equation}
where $t$ denotes the iteration index. Afterwards, we calculate the cavity distribution of $x_i$ \cite{epmimo} based on $q^{(t)}(x_i)$ as:
 \begin{equation}
 \begin{split}
q^{(t)\setminus i}(x_i)&=\frac{q^{(t)}(x_i)}{t^{(t)}(x_i)}\\
&\varpropto\frac{\mathcal{N}(x_i:\mu_i^{(t)},\Sigma^{(t)}_{ii})}{\mathcal{N}(x_i:\eta^{(t)}_iV_i^{-1^{(t)}},V_i^{-1^{(t)}})}\\
&\varpropto\mathcal{N}(x_i:m^{(t)}_{o,i},v^{(t)}_{o,i}),
\label{eq:klequalivant}
\end{split}
\end{equation}
where we have
\begin{equation}
m^{(t)}_{o,i}=\frac{\Sigma^{(t)}_{ii}}{1-\Sigma^{(t)}_{ii}V^{(t)}_i},
\label{eq:epmean2}
\end{equation}
and
\begin{equation}
v_{o,i}^{(t)}=m^{(t)}_{o,i}\left(\frac{\mu_i^{(t)}}{\Sigma^{(t)}_{ii}}-\eta^{(t)}_i\right),
\label{eq:epvar2}
\end{equation}
where $\textbf{m}^{(t)}_{o}=[m^{(t)}_{o,1},m^{(t)}_{o,2},\cdots,m^{(t)}_{o,N_tN_fU}]^T$ and the covariance vector $\textbf{V}^{(t)}_{o}=[v^{(t)}_{o,1},v^{(t)}_{o,2},\cdots,v^{(t)}_{o,N_tN_fU}]^T$  are forwarded to the \textit{estimation module}.
We can then compute the mean and variance of the approximated PDF $q(x_i)=q^{\setminus i}(x_i)p(x_i)$, where $p(x_i)$ denotes the discrete uniform distribution. Based on the moment matching method~\cite{epbayesian}, we can calculate the mean and variance of (\ref{eq:klequalivant}) as:
\begin{equation}
\hat{m}^{(t)}_{i}=\sum_{a\in\Omega}a\times q^{(t)\setminus i}(x_i=a),
\label{eq:epmean3}
\end{equation}

\begin{equation}
v_{i}^{(t)}=\sum_{a\in\Omega}(x_i-\hat{x}^{(t)})^2\times q^{(t)\setminus i}(x_i=a).
\label{eq:epvar3}
\end{equation}
The EP detector iteratively estimates the transmitted symbols and carries out hard decision at iteration $T$. The hard decision of $\hat{\textbf{x}}_{(T)}$ is based on comparing the Euclidean distances from the elements of the constellation set $\mathcal{Q}$, which is the mean of the cavity distribution $q^{(t)\setminus i}(x_i)$, as shown in (\ref{eq:epmean2}).
Furthermore, in the case of $t<T$, the parameter pair $(\eta^{(t+1)}_i,V_i^{(t+1)})$ is updated by matching the mean and variance of 
\begin{equation}
V^{(t+1)}_i=\frac{1}{v^{(t)}_{i}}-\frac{1}{v^{(t)}_{o,i}},
\label{eq:epvar4}
\end{equation}
and
\begin{equation}
\eta_i^{(t+1)}=\frac{\hat{x}_i^{(t)}}{v_i^{(t)}}-\frac{m^{(t)}_{o,i}}{v^{2(t)}_{o,i}}.
\label{eq:epmean4}
\end{equation}
The parameter update in (\ref{eq:epvar3}) may result in a negative $V_i^{(t+1)}$, which causing no matching pair of $(\eta^{(t+1)}_i,V_i^{(t+1)})$. In this case, when $V_i^{(t+1)}<0$, we have $V_i^{(t+1)}=V_i^{(t)}$ and $\eta_i^{(t+1)}=\eta_i^{(t)}$. Finally, we can harness the following damping actions
\begin{equation}
V_i^{(t+1)}=(1-\epsilon)V_i^{(t+1)}+\epsilon V_i^{(t)},
\label{eq:vardamp}
\end{equation}
and
\begin{equation}
\eta_i^{(t+1)}=(1-\epsilon)\eta_i^{(t+1)}+\epsilon \eta_i^{(t)}.
\label{eq:meandamp}
\end{equation}
where $\epsilon\in[0,1]$ is a weighting coefficient. Finally, the \textit{estimation module} sends the parameter pair to the observation module for the next iteration.
The operation of the EP detectors is summarized in \textbf{Algorithm ~\ref{alg:ep}}.
\begin{algorithm}[!t]
\DontPrintSemicolon
  
  \KwInput{$\textbf{y},\bar{\textbf{H}},\sigma^2,\Omega,\Phi$, transmit power $E_s$}
  \KwInitialization{$\hat{\textbf{x}}^{t=1}=0,\boldsymbol{\eta}^{t=1}=0,\textbf{V}^{t=1}=\frac{1}{E_s}\textbf{I}$,$\epsilon=0.05$}

  \For{$t=1,2,\cdots,T$}
   {
   \For{$i=1,2,\dots,\Omega$}
   {
       Compute $\boldsymbol{\Sigma}$ and $\boldsymbol{\mu}$ based on (\ref{eq:mean}) and (\ref{eq:variance});
       
      Compute the mean and variance of the cavity distribution $q^{(t)\setminus i}(x_i)$ based on (\ref{eq:epvar2}) and (\ref{eq:epmean2});
       
     Compute the mean and variance of the cavity distribution $q^{(t)}(x_i)$ based on (\ref{eq:epvar3}) and (\ref{eq:epmean3});
       
       Match the moment based on (\ref{eq:vardamp}) and (\ref{eq:meandamp})};

   Update the approximate mean $\textbf{m}^{(t)}$ and variance $\textbf{v}^{(t)}$;
   
   $\hat{\textbf{x}}^{(t)}=\hat{\textbf{m}}^{(t)}$\;
   }

\KwOutput{$\hat{\textbf{x}}^{(T)}$}

\caption{Expectation Propagation Detector}
\label{alg:ep}
\end{algorithm}

\subsection{GNN-based Detector}
\label{sec:gnn}
\subsubsection{GNN structure}
\label{sec:gnnstru}
GNNs offer a transformative approach to MU detection by addressing the inherent limitations of traditional detectors, such as the MMSE, AMP, and EP schemes\cite{gnnmimo}\cite{gnnreview}. Although conventional detection methods rely on predefined algorithms and assumptions, they struggle to handle high-dimensional data and hence suffer from MUI. By contrast,  GNNs exploit the structured nature of MU systems in a graph framework, dynamically learn and adapt, while improving both the detection accuracy and computational efficiency. In the next section, we explore the application of GNN-based detectors to our CS-SFIM system, including the GNN-MIMO, GNN-AMP, and GEPnet schemes. Each method demonstrates how GNNs can enhance the performance by integrating graph-based processing techniques with traditional detection frameworks. 

\begin{figure}
  \centering
  \includegraphics[width=7cm]{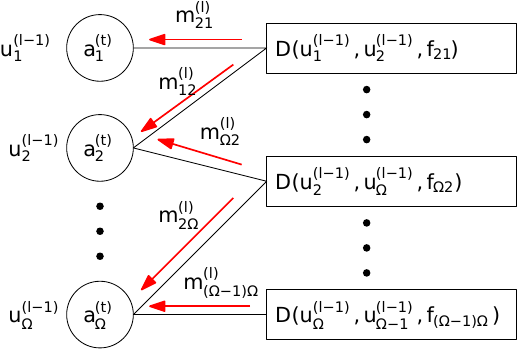}
 \caption{Propagation process of the GNN}
  \label{Figure:prop}
\end{figure}

\begin{figure}
  \centering
  \includegraphics[width=6cm]{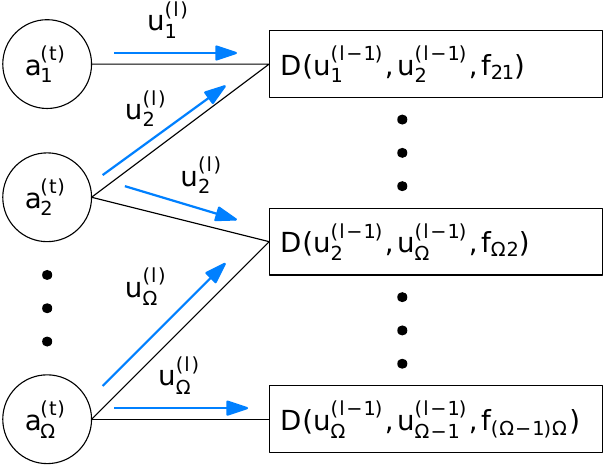}
 \caption{Aggregation process of the GNN}
  \label{Figure:agg}
\end{figure}

Since the posterior distribution approximation accuracy critically depends on the scale of the factor graph in the MP-based detector, for a limited number of nodes, we need advanced damping methods for estimating the information of each node. As shown in Fig.~\ref{Figure:fact}, we consider the TAs and RAs as the nodes and the CSI as the edges in the construction of the GNNs. Then, we can deploy the GNN framework for capturing the structured dependency of the transmitted signal by harnessing  learning method into the MP on the pair-wise MRF model as shown in Fig.~\ref{Figure:prop} and Fig.~\ref{Figure:agg}.
\begin{itemize} 
    \item \textbf{Node}: Each node $i\in\Omega$ represents the $i$-th element in the SF matrix of the equivalent symbol $\textbf{x}$.
    \item \textbf{Node attributes}: Each node is assigned its node attribute $\textbf{a}_i$ that is constant, when exchanging the information between nodes.
    \item \textbf{Edge}: An edge $e_{i,j}\in\mathcal{E}$ connects node $i\in\Omega$ and $j\in\Omega$.
    \item \textbf{Edge attributes}: Each edge $e_{i,j}$ has an assigned edge attribute $\textbf{f}_{ij}$ that is constant, when computing the message, where we use the CSI and noise as edge attribute.
     \item \textbf{Hidden vector}: Each node $\textbf{U}_i$ has a hidden vector updated in each round of the GNN and used for computing the output of the GNN.
      \item \textbf{Message}: the messages $\textbf{m}_{ji}$ arriving from the connected edges are used to update the node feature vector.
\end{itemize}
The $i$-th  variable node is characterized by a self potential $\phi(x_i)$, and the $(i,j)$-th pair of an edge is characterized by a pair potential $\psi(x_i,x_j)$, which are given by

\begin{equation}
\phi(x_i)=\exp\left(\frac{1}{\sigma^2}\textbf{y}^T\textbf{h}_ix_i-\frac{1}{2}\textbf{h}^T_i\textbf{h}_ix_i^2\right)p(x_i),
\label{eq:selfpoten}
\end{equation}
and
\begin{equation}
\psi(x_i,x_j)=\exp\left(-\frac{1}{\sigma^2}\textbf{h}^T_i\textbf{h}_ix_ix_j\right),
\label{eq:pairpoten}
\end{equation}
where $i,j\in{1,\cdots,\Omega}$ and $i\neq j$. Finally, the joint probability $p_{out}(\textbf{x})$ corresponding to the pair-wise MRF can be obtained as \cite{gnnmimo}
\begin{equation}
p_{out}(\textbf{x})=\frac{1}{Z}\prod_{i=1}^{\Omega}\phi(x_i)\prod_{j=1 j\neq i}^{\Omega}\psi(x_i,x_j),
\label{eq:pairwiseMRF}
\end{equation}
where $Z$ is a normalization constant.\par
 GNN  can be used to infer the \textit{a posteriori} probability  $p(\textbf{x}|\textbf{y})$ and to recover the equivalent transmitted symbols $\textbf{x}$ in the CS-SFIM detection problem of (\ref{equation:alluserdet2}). 
In this case, GNNs are constructed for the MIMO pair-wise MRF presented in  (\ref{eq:selfpoten}) and (\ref{eq:pairpoten}). The MRF models the structured dependency of a set of random variables $\textbf{x}={x_1,\cdots,x_i,\cdots,x_{\Omega}}$ by an undirected graph $G=\{\Omega,\mathcal{E}\}$, where $\Omega$ and $\mathcal{E}$ are the set of nodes and edges, respectively. \par
Then, the input of the GNN is extracted from $\phi(x_i)$ and $\psi(x_i,x_j)$ and the information along each edge $e_{i,j}$ represents the feature vector $\textbf{f}_{ij} = [\textbf{h}_i^T\textbf{h}_j, \sigma^2]$, which can be obtained from (\ref{eq:pairwiseMRF}) and utilized for the message passing of the GNN. The hidden vector $\textbf{u}_i^{0}$ of each node $i$ is initialized with 
\begin{equation}
\textbf{u}_i^0=\textbf{W}_1[\textbf{y}^T \textbf{h}_i,\textbf{h}_i^T\textbf{h}_i,\sigma^2]^T+\textbf{b}_1,
\label{eq:gnninitunit}
\end{equation}
where we have $\textbf{W}\in\mathbb{R}_{N_u}\times \mathbb{R}^3$ and $\textbf{b}\in\mathbb{R}^{N_u}$ is a learnable vector, while $N_u$ is the size of the feature vector. 
A typical GNN framework consists of three main modules: propagation module, aggregation module and readout module. The first two modules operate at every iteration $l$,  while the readout module is activated only at final iteration $L$ to make the inference. In the following we elaborate on the GNN processing.

\paragraph{Propagation module}
Fig.~\ref{Figure:prop} shows a typical GNN structure  and its  propagation process. The messages $u_k^{(l)}, k\in [\Omega],l\in[L]$ are sent to the neighboring factor nodes and the propagation module will gather the messages from pairs of variable nodes, and feed back the update message for each direct edge. For example, the $m_{21}^{(l)}$ is the update message of edge $e_{21}$ in the $l$ iteration and fed back to node 1. 
The propagation module outputs the updated message $\textbf{m}^{(i)}_{ji} $ for each direct edge $e_{ij}\in\mathcal{E}$ as:

\begin{equation}
\textbf{m}^{(l)}_{ji}=\mathcal{D}(\textbf{u}_i^{(l-1)},\textbf{u}_j^{(l-1)},\textbf{f}_{ji}),
\label{eq:gnnmessage}
\end{equation}
where $\textbf{f}_{ji}$ is the information associated with the edge $e_{ij}$ and  $\mathcal{D}$ is a Multiple Layer Perceptron (MLP) using the classic Rectified Linear Unit (ReLU) as activation function. In the propagation module, each edge has an MLP associated with two hidden layers of size $N_{h_1}$ and $N_{h_2}$ and an output layer of size $N_u$.  Finally, the outputs $\textbf{m}^{(l)}_{ji}$ are fed back to the nodes.

\paragraph{Aggregation module}
The aggregation module of each variable node adds all the incoming messages $m_{jk}^{(l)},j,i\in[\Omega]$ gleaned from its connected edges and concatenates the sum of $\textbf{m}_{ji}^{(l)}$ with the node attribute $\textbf{a}_i^{(t)}$ to obtain the gathering message $m_i{(l)}$ for the current iteration $l$, which is used to update the node message $u^{(l)}_i$ to the propagation module for the next iteration, as shown in Fig.~\ref{Figure:agg}:

\begin{equation}
\textbf{m}_{i}^{(l)}=\left[\sum^{\Omega}_{j=1, j\neq i}\textbf{m}_{ji}^{(l)},\textbf{a}_n^{(t)}\right].
\label{eq:gnnnodesum}
\end{equation}
 The message $\textbf{m}_{i}^{(l)}$ is used for computing the nodes' hidden vector $\textbf{u}_i^{(l)}$ as follow:

\begin{equation}
\textbf{g}_i^{(l)}= U(\textbf{g}_i^{(l-1)},\textbf{m}_i^{(l)}),
\label{eq:gnnhidden1}
\end{equation}
\begin{equation}
\textbf{u}_i^{(l)}= \textbf{W}_2\textbf{g}_i^{(l)}+\textbf{b}_2,
\label{eq:gnnhidden2}
\end{equation}
where the function $U$ is constituted by a Gated Recurrent Unit (GRU) \cite{gru} based network associated with the current and the previous states. The previous hidden states specified as $\textbf{g}_i^{(l)}\in\mathbb{R}^{N_{h_1}}$ and $\textbf{g}_i^{(l-1)}\in\mathbb{R}^{N_{h_1}}$, respectively. Furthermore, $\textbf{W}_2\in\mathbb{R}^{N_{u}\times N_{h_1}}$ is a learnable matrix and $\textbf{b}_2\in\mathbb{R}^{N_{u}}$ is a learnable vector. The updated hidden vector in (\ref{eq:gnnhidden2}) is passed to the propagation module for the next iteration, as shown in Fig.~\ref{Figure:agg}.

\begin{figure}
  \centering
  \includegraphics[width=5cm]{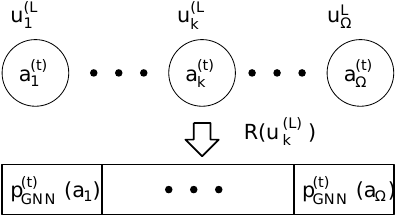}
 \caption{Readout process of the GNN}
  \label{Figure:read}
\end{figure}

\paragraph{Readout module}
In Fig.~\ref{Figure:read}, after $L$ iterations of the MP between the propagation and aggregation modules, a readout module is harnessed for generating the estimated distribution $p_{GNN}^{(t)}(x_i=s_i|\textbf{y})$ for the $t$-layer GNN given by 
\begin{equation}
\Tilde{p}_{GNN}^{(t)}(x_i=a|\textbf{y})=R(\textbf{u}_i^{(l)}), a\in\Omega,
\label{eq:gnnread1}
\end{equation}
and
\begin{equation}
p_{GNN}^{(t)}(x_i=a|\textbf{y})=\frac{\exp{(\Tilde{p}_{GNN}^{(t)}(x_i=a|\textbf{y}))}}{\sum_{a\in\Omega}\exp{(\Tilde{p}_{GNN}^{(t)}(x_i=a|\textbf{y})})}, a\in\Omega.
\label{eq:gnnread2}
\end{equation}
The readout function $R$ consists of an MLP having two hidden layers of sizes $N_{h_1}$ and $N_{h_2}$ along with the classic RELU activation after each hidden layer. Hence, we can compute
\begin{equation}
\textbf{g}_i^{(0)}\leftarrow \textbf{g}_i^{(L)} \; \text{and} \; \textbf{u}_i^{(0)}\leftarrow \textbf{u}_i^{(L)}, i=1,2,\cdots,\Omega.
\label{eq:iteration}
\end{equation}
for the next iteration. 

\subsubsection{GNN-MMSE}
In this section, to improve the prior information, we incorporate the MMSE \textit{a posteriori} as the \textit{a priori} $p(x)$ derived from (\ref{equation:alluserdet2}) such that
\begin{equation}
p(x_i)=\frac{1}{\pi h_{c_{ii}}}\exp{[-\frac{(z_i-x_i)^2}{h_{c_{ii}}}}],
\label{eq:gnnmmsemimo}
\end{equation}
where $z_i$ is the $i$-th element of the MMSE estimation vector $\textbf{z}=(\bar{\textbf{H}}^T\bar{\textbf{H}}+\sigma^2\textbf{I}_{\Omega})^{-1}\bar{\textbf{H}}^T\textbf{y}$ and $\sigma_{h_{c_{ii}}}$ is the $(i,i)$-th element of $\boldsymbol{C}=\sigma^2(\bar{\textbf{H}}^T\bar{\textbf{H}}+\sigma^2\textbf{I}_{\Omega})^{-1}$. The prior correlation coefficient $\rho_{ij}$ between the variable $x_i$ and $x_j$, namely $\rho_{ij}=\frac{h_{c_{ij}}^2}{h_{c_{ii}}h_{c_{jj}}}$ is added to the feature vector $\textbf{f}_{ij}$.
Afterwards, the same model of Section \ref{sec:gnnstru} is applied subject to the slight modification of the information  $\textbf{f}_{ij}$ between the edges and the initial value of the hidden vector $\textbf{u}_i^{(0)}$. Then, the information along the edge $e_{ij}$ is defined as $e_{ij}=[\rho_{ij}, \textbf{h}_i^T\textbf{h}_i,\sigma^2]$ and the initial hidden vector $\textbf{u}_i^{(0)}$ of each node $i$ is initialized with $\textbf{u}_i^{(0)}=\textbf{W}_1[z_i,h_{c_{ii}},\textbf{y}^T\textbf{h}_i,\textbf{h}_i^T\textbf{h}_i,\sigma^2]^T+\textbf{b}_1$.
\subsubsection{AMP-GNN}
\begin{figure}
  \centering
  \includegraphics[width=8.5cm]{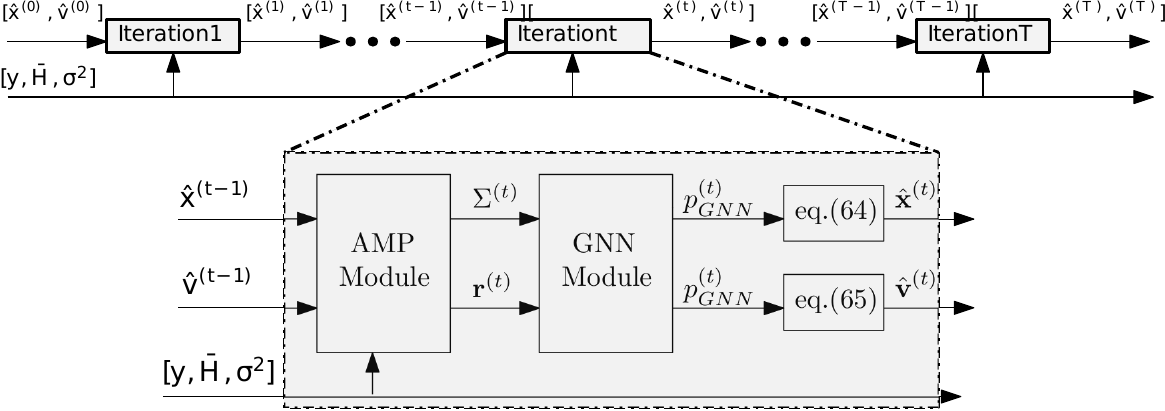}
 \caption{The structure of the AMP-GNN network.}
  \label{Figure:ampgnn}
\end{figure}

The GNN can also be harnessed for the MP-based method to reduce the inaccuracy of approximation. In this case, the conventional AMP approximates the distribution of information by the Gaussian distribution, but the GNN can be invoked for achieving a more accurate distribution of the corresponding information.
As shown in Fig.~\ref{Figure:ampgnn}, in the $t$-th iteration, the AMP-GNN takes its input from the linear module in the AMP algorithm and then the mean $r_i^{(t)}$ and variance $\Sigma_i^{(t)}$ from (\ref{eq:amp3}) and (\ref{eq:amp4}) are fed into the twin-tuple attribute $\textbf{a}_n^t$ of the variable node $x_i$ as follows:
\begin{equation}
\textbf{a}_n^t=[r_i^{(t)},\Sigma_i^{(t)}].
\label{eq:gnnamp}
\end{equation}

After a few iterations, we can invoke $p_{GNN}^{(t)}(x_i=s_i|\textbf{y})$ of Fig.~\ref{Figure:ampgnn} to compute the \textit{a posteriori} mean and variance  for the AMP, formulated as: 
\begin{equation}
   \hat{x}^{(t+1)}_i=\mathbb{E}(x_i;p_{GNN}^{(t)}),
    \label{eq:ampgnnmean}
\end{equation}
and
\begin{equation}
    \hat{v}^{(t+1)}_i=Var(x_i;p_{GNN}^{(t)}),
    \label{eq:amgnnpvar}
\end{equation}
where the expectation and variance of the estimated signal $x_i$ are computed by taking into account $p_{GNN}^{(t)}$. The most significant difference between conventional AMP and AMP-GNN is that $p_{GNN}^{(t)}$ is not approximated by the Gaussian distribution for the AMP-GNN detector. Then, $\hat{x}^{(t+1)}_i$ and $\hat{v}^{(t+1)}_i$ are forwarded to the next AMP-GNN iteration, where the iterations terminate at a fixed number of layers. 
Accordingly, we can modify \textbf{Algorithm~\ref{alg:amp}} leading to the AMP-GNN procedure of \textbf{Algorithm~\ref{alg:ampgnn}}.
\begin{algorithm}[!ht]
\DontPrintSemicolon
  
  \KwInput{$\textbf{y},\bar{\textbf{H}},\sigma^2,p(x_i)$}
  \KwInitialization{$\hat{x}^{t=1}_i=0,\hat{v}^{t=1}_i=1,Z^{t=0}_j=y_j,\textbf{g}^{(0)}=0$}
  
  \For{$t=1,2,\cdots,T$}
   {
    Compute $\Sigma^{(t)}_i$ and $r^{(t)}_i$ in (\ref{eq:amp3}) and (\ref{eq:amp4}), $i\in[\Omega]$\;

    \textbf{The GNN Module}\;
    
    Compute (\ref{eq:gnnamp})\;

    \If{$t=1$}
    {
        Compute $\textbf{u}_i^{(0)}$ in (\ref{eq:gnninitunit}), $i\in[\Omega]$\;
    }

    \For{$l=1,2,\dots,L$}
    {    
        Compute $\textbf{m}_{ji}^{(l)}$ and $\textbf{m}_{i}^{(l)}$ in (\ref{eq:gnnmessage}) and (\ref{eq:gnnnodesum}), $i,j\in[\Omega], i\neq j$\;
        
        Compute $\textbf{g}_{i}^{(l)}$ and $\textbf{u}_{i}^{(l)}$ in (\ref{eq:gnnhidden1}) and (\ref{eq:gnnhidden2}), $i\in[\Omega]$\;
    } 
    
    Compute $p_{GNN}^{(l)}(x_i)$ in (\ref{eq:gnnread1}) and (\ref{eq:gnnread2})\;
    
    Compute (\ref{eq:iteration})\;

    Compute $\hat{x}^{(t+1)}_i$ and $\hat{v}^{(t+1)}_i$ based on (\ref{eq:ampgnnmean}) and (\ref{eq:amgnnpvar}), $i\in[\Omega]$\;

     }
\KwOutput{$\hat{\textbf{x}}^{(T)}$}

\caption{AMP-GNN Detector}
\label{alg:ampgnn}
\end{algorithm}
\subsubsection{GEPnet}
To mitigate the inaccuracy of the posterior distribution approximation in the EP detector,  we harness a feature proposed in \cite{gnnepmimo}, which treats the EP's cavity $q^{(t)\setminus i}(x_i)$ in (\ref{eq:klequalivant}) as the prior information for the variable node $i$ in our GEPnet detector. The cavity $q^{(t)\setminus i}(x_i)$ is given by a Gaussian distribution with a mean $m^{(t)}_{o,i}$ and variance $v_{o,i}^{(t)}$ defined in (\ref{eq:epmean2}) and (\ref{eq:epvar2}). Then, the variable node attribute can be represented as:

\begin{equation}
\textbf{a}_n^t=[m^{(t)}_{o,i},v_{o,i}^{(t)}].
\label{eq:gepattrib}
\end{equation}

\begin{figure}
  \centering
  \includegraphics[width=8.5cm]{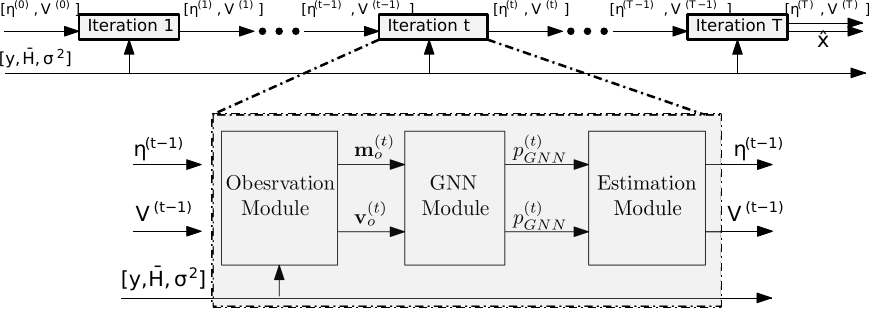}
 \caption{The structure of the GEPnet.}
  \label{Figure:gep}
\end{figure}

Hence, we can modify the GEPnet readout module of (\ref{eq:gnnread2}), where the appropriately adjusted cavity distribution $q^{(t)\setminus i}_{GNN}(x_i=a)$ can be represented as
\begin{equation}
q^{(t)\setminus i}_{GNN}(x_i=a)=\frac{\exp{[\Tilde{p}_{GNN}^{(t)}(x_i=a)]}}{\sum_{a\in\Omega}\exp{[\Tilde{p}_{GNN}^{(t)}(x_i=a)]}}, a\in\Omega.
\label{eq:gnnepread}
\end{equation}
Then, we can use the GNN modules, trained distribution  for replacing the distribution in the \textit{estimation module} of the EP as  
\begin{equation}
\hat{m}^{(t)}_{i}=\sum_{a\in\Omega}a\times q^{(t)\setminus i}_{GNN}(x_i=a),
\label{eq:gepestmean}
\end{equation}

\begin{equation}
v_{i}^{(t)}=\sum_{a\in\Omega}(x_i-\hat{x}^{(t)})^2\times q^{(t)\setminus i}_{GNN}(x_i=a).
\label{eq:gepestvar}
\end{equation}
The GEPnet structure is shown in Fig.~\ref{Figure:gep} and the GEPnet procedure is presented in \textbf{Algorithm~\ref{alg:gep}}.

\begin{algorithm}[!htb]
\DontPrintSemicolon
  
  \KwInput{$\textbf{y},\bar{\textbf{H}},\sigma^2,\Omega,\Phi$, transmit power $E_s$}
  \KwInitialization{$\hat{\textbf{x}}^{t=1}=0,\boldsymbol{\eta}^{t=1}=0,\textbf{V}^{t=1}=\frac{1}{E_s}\textbf{I}$,$\epsilon=0.1,\textbf{g}^{(0)}=0$}
 \KwIteration{}
 \For{$t=1,2,\cdots,T$}
   {
    \textbf{EP Observation Module}\;
    
    Compute $\boldsymbol{\Sigma}$ and $\boldsymbol{\mu}$ based on (\ref{eq:mean}) and (\ref{eq:variance})\;
    
    Compute $m^{(t)}_{o,i}$ and $v_{o,i}^{(t)}$ based on (\ref{eq:epmean2}) and (\ref{eq:epvar2})\;
    
    Compute (\ref{eq:gepattrib})\;
    
    \textbf{The GNN Module}\;

    \If{$t=1$}
    {
        Compute $\textbf{u}_i^{(0)}$ in (\ref{eq:gnninitunit}), $i\in[\Omega]$\;
    }

    \For{$l=1,2,\dots,L$}
    {    
        Compute $\textbf{m}_{ji}^{(l)}$ and $\textbf{m}_{i}^{(l)}$ in (\ref{eq:gnnmessage}) and (\ref{eq:gnnnodesum}), $i,j\in[\Omega], i\neq j$\;
        
        Compute $\textbf{g}_{i}^{(l)}$ and $\textbf{u}_{i}^{(l)}$ in (\ref{eq:gnnhidden1}) and (\ref{eq:gnnhidden2}), $i\in[\Omega]$\;
    } 
    
    Compute $q_{GNN}^{(l)}(x_i)$ in (\ref{eq:gnnepread})\;
    
    Compute (\ref{eq:iteration})\;

    \textbf{EP Estimation Module}\;
    
    Compute $\hat{m}^{(t)}_{i}$ and $v_{i}^{(t)}$ based on (\ref{eq:epmean3}) and (\ref{eq:epvar3}), $i\in[\Omega]$\;
    
    Compute $\eta_i^{(t+1)}$ and $V_i^{(t+1)}$ based on (\ref{eq:epvar4}) and (\ref{eq:epmean4}), $i\in[\Omega]$\;
    
    Optimize $\eta_i^{(t+1)}$ and $V_i^{(t+1)}$ based on (\ref{eq:vardamp}) and (\ref{eq:meandamp})\;

   }
   $\hat{\textbf{x}}^{(t)}=\hat{\textbf{m}}^{(t)}$;\

\KwOut{$\hat{\textbf{x}}^{(T)}$}

\caption{GEPnet Detector}
\label{alg:gep}
\end{algorithm}

\subsection{Complexity analysis}
In this section, we analyze the computational complexity of the GNN-based detector of our CS-SFIM system, then compare it to conventional MP-based MU detectors and classic DNN-based detector, which use results of ML detection as training target. The complexity of the AMP detector is on the order of $\mathcal{O}(\Omega\Phi)$ due to the associated matrix-vector multiplication and the complexity of the EP is $\mathcal{O}(\Omega^2\Phi)$. Additionally, the complexity of the DNN\cite{xinyudnn}, GNN-MMSE, AMP-GNN and GEPnet is $\mathcal{O}(N_{h_1}N_{h_2})$, $\mathcal{O}(\Omega^2+\Omega^2 F^2)$, $\mathcal{O}(\Omega\Phi+\Omega^2 F^2)$ and $\mathcal{O}(\Omega^2\Phi+\Omega^2 F^2)$.\footnote{$h_1$ and $h_2$ represent the input and output of the DNN. $F$ represents the features of each node in GNN network.}
Then we can characterize the complexity of each detector in different scenarios in Table~\ref{Table:total} and computational complexity based on the number of multiplications as our metric quantified in Table~\ref{Table:complex}.

\begin{table*}
  \centering
  \caption{Multiplications number of different detectors for MU-CS-SFIM.}
  \begin{tabular}{l|c|c|c|c|c|c|c|c|c} 
  \hline
 \multicolumn{3}{c|}{MU setting}&MMSE&AMP&EP&DNN&GNN-MMSE&AMP-GNN&GEPnet\\
  \hline
   User&TA&RA&\multicolumn{6}{c}{}\\
  \hline
  4&4&16&$4.31\times10^{6}$&$1.12\times10^{6}$&$2.61\times10^{7}$&$2.43\times 10^{6}$&$2.13\times10^{7}$&$2.75\times10^{7}$&$6.72\times10^{7}$\\
  \hline
  16&4&64&$1.42\times10^{7}$&$4.35\times10^{6}$&$5.23\times10^{8}$&$9.83\times 10^{6}$&$9.22\times10^{7}$&$1.33\times10^{8}$&$8.84\times10^{8}$\\
  \hline
16&4&128&$1.42\times10^{7}$&$8.71\times10^{6}$&$1.12\times10^{9}$&$1.05\times 10^{8}$&$1.15\times10^{8}$&$1.54\times10^{8}$&$1.25\times10^{9}$\\
  \hline
  \end{tabular}
  \label{Table:complex}
\end{table*}
\section{Performance Analysis}
\label{sec:perf}
In this section, we first characterize the MU {CS-SFIM} system proposed in Section~\ref{sec:system}.  We then compare the performance of the GNN-based detectors to that of conventional detectors. More specifically, the BER is evaluated both as a function of the number of users and of the transmit-to-receive antenna ratio.  We consider $U=4$ and $16$ users. For $U=4$ users, we employ 4 TAs per user and 16 RAs. Hence, the transmit-to-receive antenna ratio is $\varrho\equiv\frac{N_t\cdot U}{N_r}=1$. By contrast, we set $\varrho\in[\frac{1}{2},1]$ for the $U=16$ users' scheme.   We investigated a set of schemes, which are summarized as follows:

  \begin{enumerate}

 \item \textbf{Scheme 1}: MU-{CS}-SFIM, which supports $U=4$ users, activating $N_a=2$ out of 4 TAs, 16 RAs, APM constellation  $\mathcal{L}=2$ and $N_f=2$ subcarriers per group, while considering $N_v=4$ subcarriers per group in the VD and $K=2$ subcarriers activated. Then we apply the  following detectors:
\begin{enumerate}
  \item MMSE based detector.
  \item AMP detector.
  \item EP detector.
    \item DNN detector.
   \item GNN-MMSE detector
  \item AMP-GNN detector.
  \item GEPnet detector.
  \item ML detector.
  \end{enumerate}

 \item \textbf{Scheme 2}: MU-{CS}-SFIM, which supports $U=4$ users, activating $N_a=2$ out of 4 TAs, 16 RAs, APM constellation  $\mathcal{L}=4$ and $N_f=2$ subcarriers per group, while considering $N_v=4$ subcarriers per group in the VD and $K=2$ subcarriers activated. Then we apply the following detectors:
\begin{enumerate}
  \item MMSE based detector.
  \item AMP detector.
  \item EP detector.
   \item DNN detector.
   \item GNN-MMSE detector
  \item AMP-GNN detector.
  \item GEPnet detector.
  \item ML detector.
  \end{enumerate}

\item \textbf{Scheme 3}: MU-{CS}-SFIM, which supports $U=16$ users, activated $N_a=2$ out of 4 TAs, 64 RAs, APM constellation  $\mathcal{L}=2$ and $N_f=2$ subcarriers per group, while considering $N_v=4$ subcarriers per group in the VD and $K=2$ subcarriers activated. Then we apply the following detectors:
\begin{enumerate}
  \item MMSE based detector.
  \item AMP detector.
  \item EP detector.
   \item GNN-MMSE detector
  \item AMP-GNN detector.
  \item GEPnet detector.
  \end{enumerate}

\item \textbf{Scheme 4}: MU-{CS}-SFIM, which supports $U=16$ users, activated $N_a=2$ out of 4 TAs, 128 RAs, APM constellation  $\mathcal{L}=2$ and $N_f=2$ subcarriers per group, while considering $N_v=4$ subcarriers per group in the VD and $K=2$ subcarriers activated. Then we apply the following detectors:
  \begin{enumerate}
  \item MMSE based detector.
  \item AMP detector.
  \item EP detector.
  \item GNN-MMSE detector
  \item AMP-GNN detector.
  \item GEPnet detector.
  \end{enumerate}

\item \textbf{Scheme 5}: MU-{CS}-SFIM, which supports $U=4$ users, where $K_a=3$ only 3 out of 4 users randomly selected are activated each time,  $N_a=2$ out of 4 TAs, 64 RAs, APM constellation $\mathcal{L}=2$ and $N_f=2$ subcarriers per group, while considering $N_v=4$ subcarriers per group in the VD and $K=2$ subcarriers activated. Then we apply the  following detectors:
  \begin{enumerate}
  \item MMSE based detector.
  \item AMP detector.
  \item EP detector.
  \item DNN detector.
    \item GNN-MMSE detector
  \item AMP-GNN detector.
  \item GEPnet detector.
  \item ML detector.
  \end{enumerate}
 
  \end{enumerate}

\begin{figure}[!htb]
\centering
\subfigure[Performance comparison of detectors in Scheme 1. ]{                    
\begin{minipage}{8.8cm}
\includegraphics[width=8.8cm]{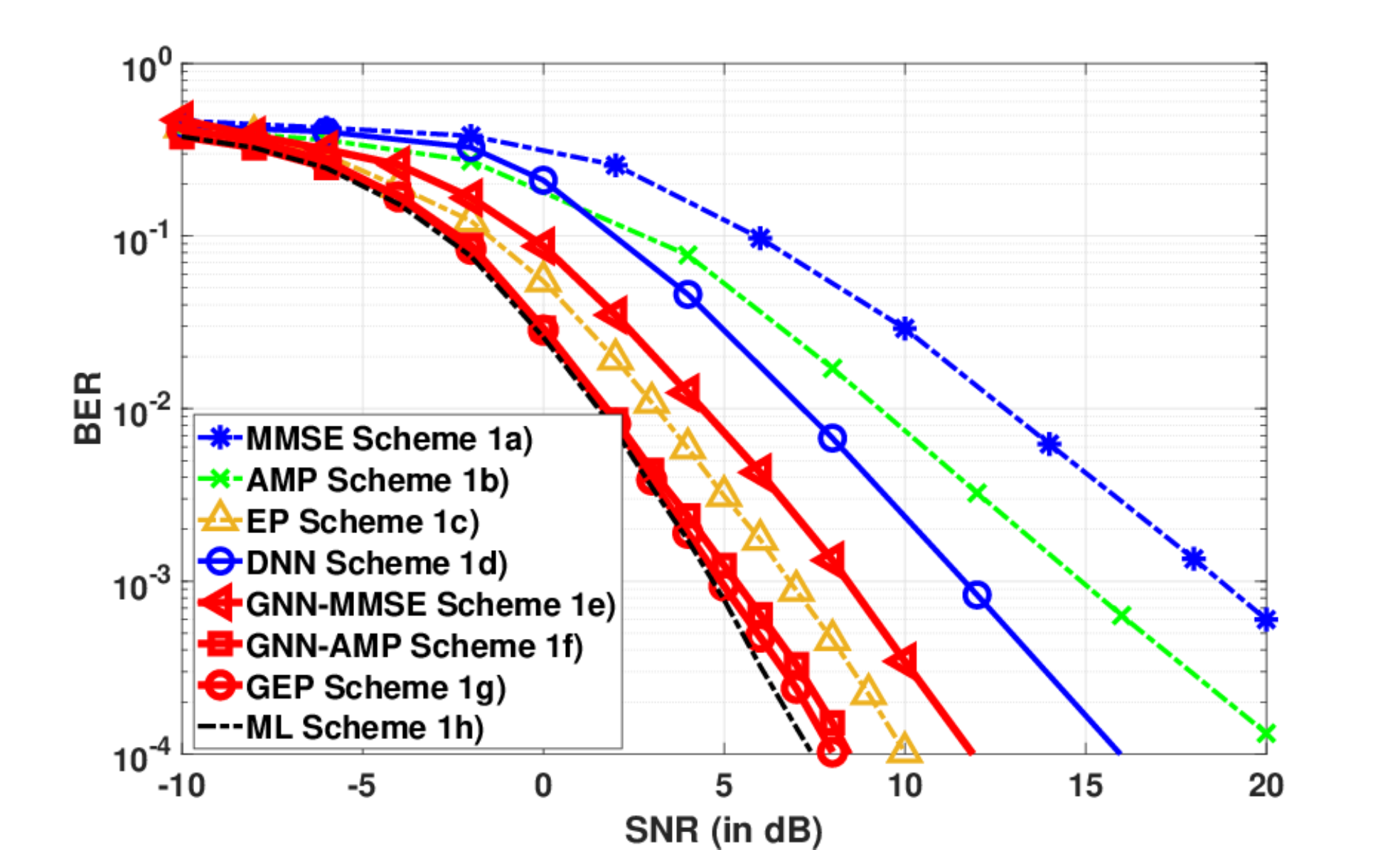}  
\label{Figure:s1}
\end{minipage}}
\subfigure[Performance comparison of detectors in Scheme 2.]{                    
\begin{minipage}{8.8cm}
\includegraphics[width=8.8cm]{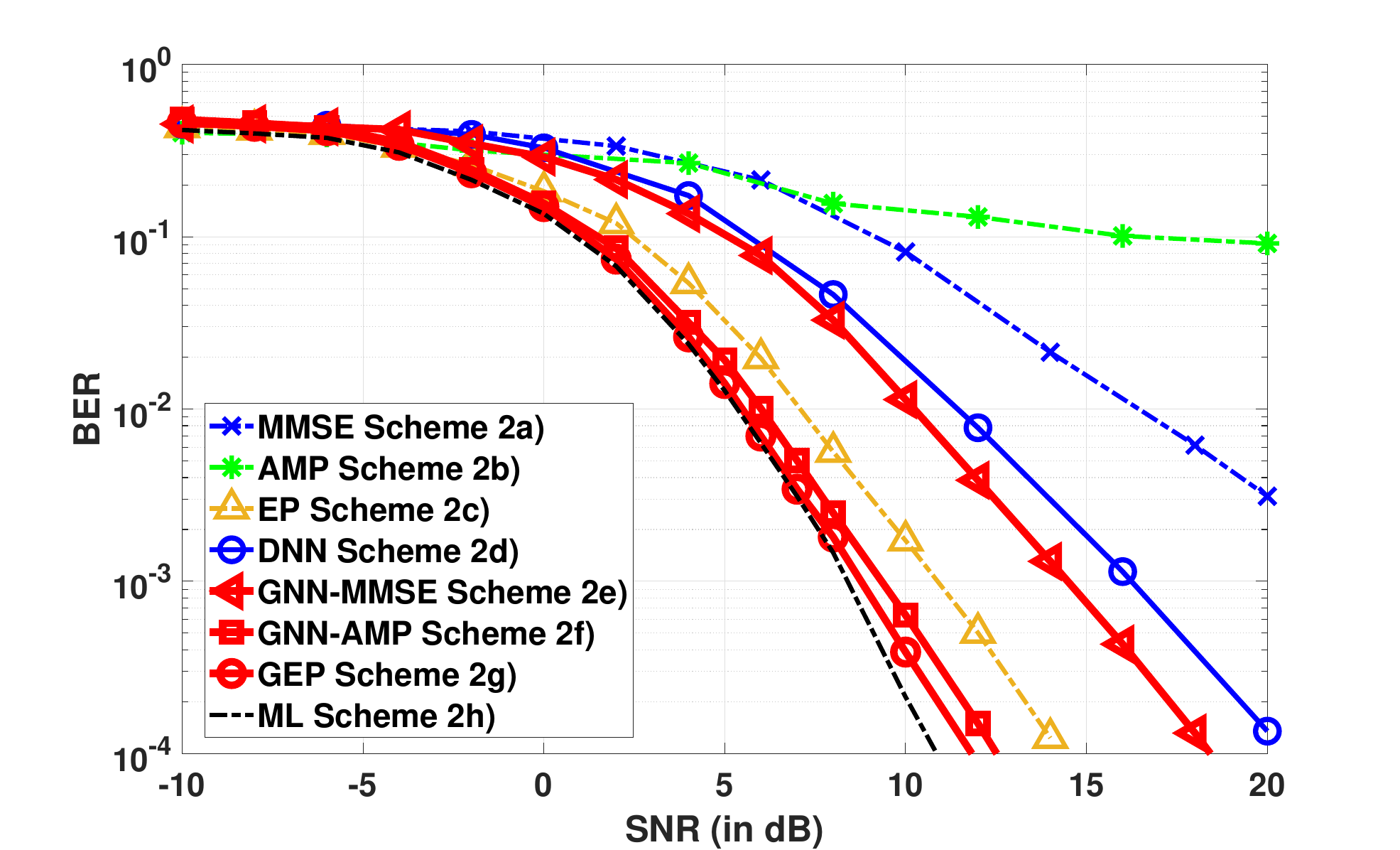} 
\label{Figure:s2}
\end{minipage}}
\caption{{BER} performance comparison of the different MU CS-SFIM schemes under different QAM schemes for transmission over Rayleigh channel.}   
\label{fig:sa1}  
\centering
\end{figure}

\begin{figure}[!htb]
\centering
\subfigure[Performance comparison of detectors in Scheme 3. ]{                    
\begin{minipage}{8.8cm}
\includegraphics[width=8.8cm]{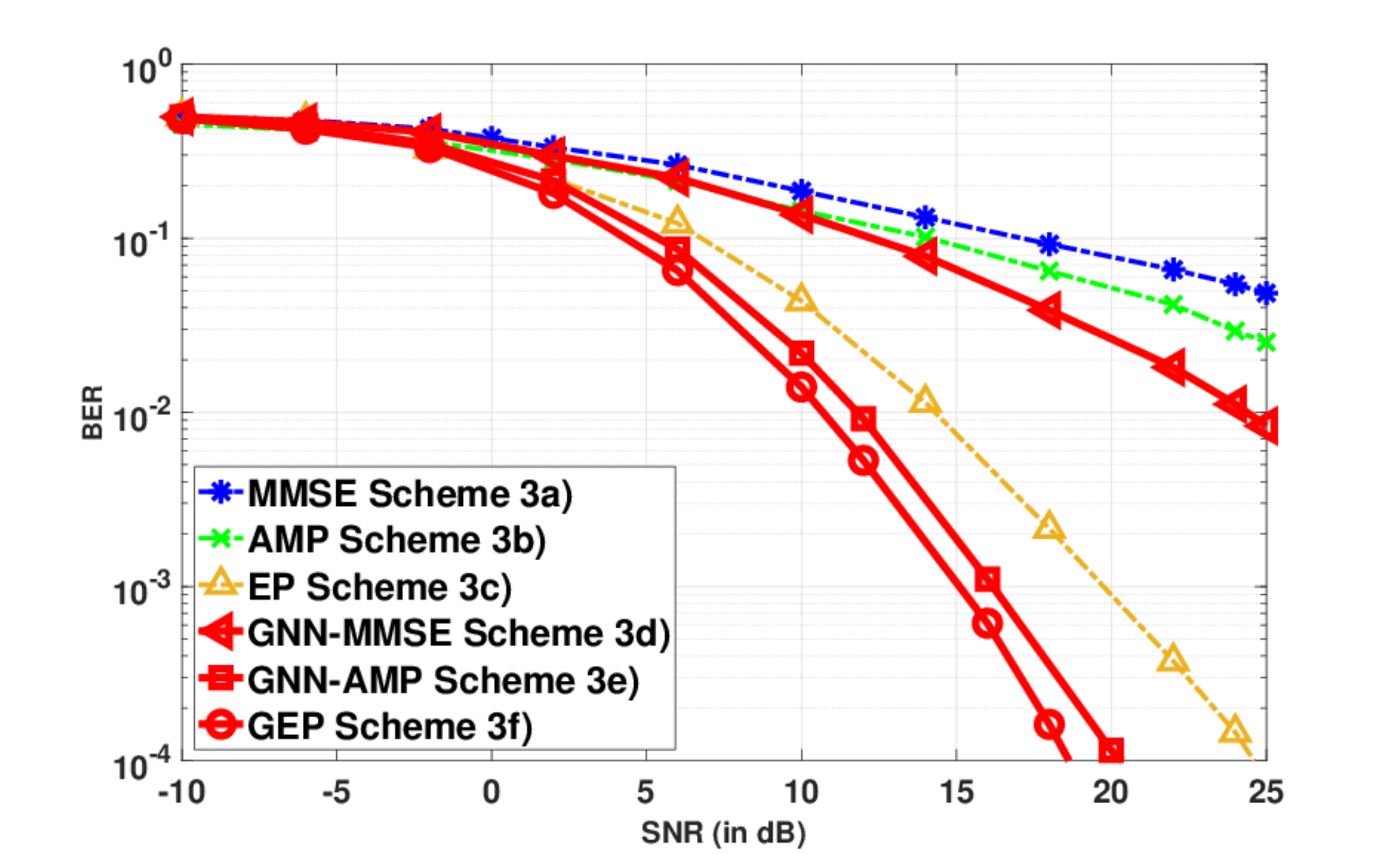}  
\label{Figure:s3}
\end{minipage}}
\subfigure[Performance comparison of detectors in Scheme 4.]{                    
\begin{minipage}{8.8cm}
\includegraphics[width=8.8cm]{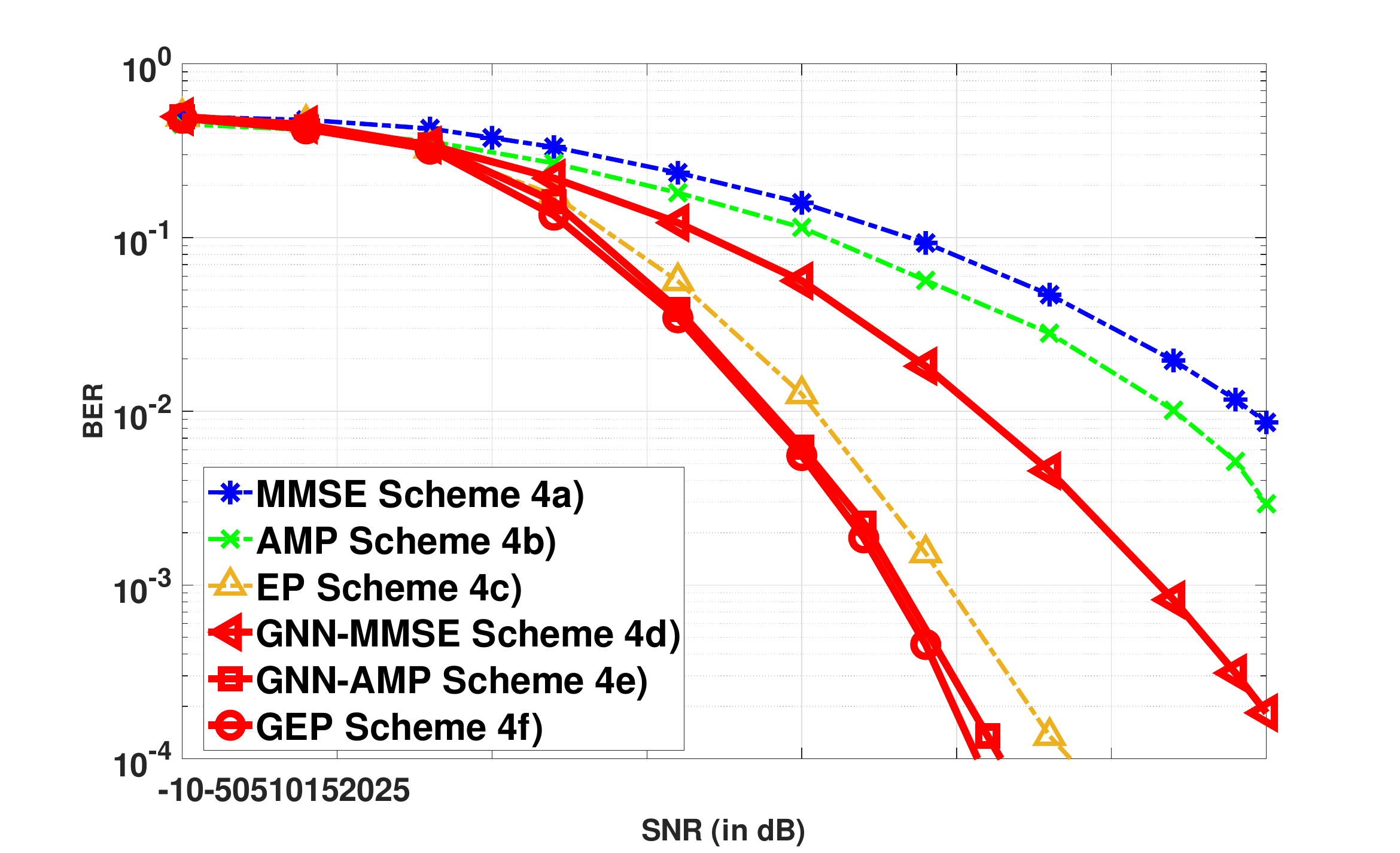} 
\label{Figure:s4}
\end{minipage}}

\caption{{BER} performance comparison of the different MU CS-SFIM schemes under different $U$ for transmission over Rayleigh channel.}   
\label{fig:sa2}  
\centering
\end{figure}

\begin{table}[!tb]
  \centering
  \caption{The SNR required by each Scheme at BER$=10^{-2}$.}
 \scalebox{0.92}{
  \begin{tabular}{|l|c|c|c|} 
  \hline
   \multicolumn{2}{|c|}{Scheme index}&SNR at BER of $10^{-2}$&Computational Complexity $\mathcal{O}(.)$\\
  \hline
  \multirow{7}{*}{Scheme 1} &{a)}&12.8&$1.6\times10^{5}$\\
    \cline{2-4}
   
   &{b)}&9.47&$6\times10^{3}$\\
   \cline{2-4}
  &{c)}&3.12&$3.8\times10^{5}$\\ 
   \cline{2-4}
    &{d)}&7.44&$4.1\times10^{3}$\\ 
   \cline{2-4}
   &{e)}&4.41&$3.2\times10^{5}$\\
   \cline{2-4}
   &{f)}&\textbf{1.77}&$3.5\times10^{5}$\\
   \cline{2-4}
   &{g)}&\textbf{1.75}&$1.6\times10^{6}$\\
   \cline{2-4}
    &{h)}&1.73&$(4.1\times10^{3})^{4}$\\

     \hline
 \multirow{6}{*}{Scheme 2}&{a)}&17.13&$2.56\times10^{6}$\\
     \cline{2-4}
    &{b)}&-&$2.4\times10^{4}$\\
    \cline{2-4}
   &{c)}&6.71&$1.5\times10^{6}$\\
   \cline{2-4}
   &{d)}&3.59&$1.3\times10^{4}$\\
   \cline{2-4}
  &{e)}&10.05&$3.5\times10^{6}$\\
   \cline{2-4}
   &{f)}&\textbf{5.33}&$2.3\times10^{6}$\\
   \cline{2-4}
   &{g)}&\textbf{5.25}&$4.6\times10^{6}$\\
   \cline{2-4}
    &{h)}&5.13&$(6.6\times10^{4})^4$\\
   \cline{2-4}

  \hline
    \multirow{6}{*}{Scheme 3}&{a)}&37.31&$6.6\times10^{5}$\\
     \cline{2-4}
    &{b)}&32.29&$3.3\times10^{4}$\\
    \cline{2-4}
   &{c)}&14.23&$8.4\times10^{6}$\\
   \cline{2-4}
  &{d)}&24.37&$1.3\times10^{5}$\\
   \cline{2-4}
   &{e)}&\textbf{11.75}&$1.7\times10^{5}$\\
   \cline{2-4}
   &{f)}&\textbf{10.72}&$8.9\times10^{6}$\\
   \cline{2-4}
  \hline

    \multirow{6}{*}{Scheme 4}&{a)}&24.51&$6.6\times10^{5}$\\
     \cline{2-4}
    &{b)}&22.04&$6.6\times10^{4}$\\
    \cline{2-4}
   &{c)}&10.41&$1.7\times10^{5}$\\
   \cline{2-4}
  &{d)}&15.72&$5.2\times10^{5}$\\
   \cline{2-4}
   &{e)}&\textbf{8.92}&$2.4\times10^{5}$\\
   \cline{2-4}
   &{f)}&\textbf{8.71}&$9.2\times10^{6}$\\
   \cline{2-4}
  \hline

  \end{tabular}
  }
  \label{Table:total}
\end{table}
Graph-based modeling naturally matches the factor-graph interpretation of the joint detection problem in CS and IM, enabling direct incorporation of the sparse prior and Markov structure~\cite{gnnmimo}.
DNNs do not inherently exploit the sparse factor-graph structure, and might need more parameters to learn an equivalent message-passing approach. 

In Fig.~\ref{Figure:s1},  we compare the performance of various detectors for our MU CS-SFIM system that supports $U=4$ users in \textbf{ scheme 1}. As discussed in Section \ref{sec:det}, for every subcarrier group of each user, we have $N_{SI}$ possible active subcarrier index combinations, $N_{AC}^K$ possible active TA index realizations, and $L^{K\cdot N_{a}}$ possible classical APM symbols combinations. The complexity order of \textbf{Scheme 1a)} is $\mathcal{O}_{MMSE}[{U}\cdot(N_{FI}N_{AC}^{K}L^{KN_a})]$ and of \textbf{Scheme 1g)} is $\mathcal{O}_{ML}[(N_{FI}N_{AC}^{K}L^{KN_a})^{U}]$, which is feasible for a small number of users. \textbf{Scheme 1g)} achieves a BER of $10^{-4}$ at an SNR of 7.12 dB, as shown in Fig.~\ref{Figure:s1}.  \textbf{Scheme 1a)} shows significant performance degradation compared to \textbf{Scheme 1g)}, since it requires 20 dB SNR to achieve a BER of $10^{-4}$. The \textbf{Scheme 1b)}, which had knowledge of \textit{a priori} information of the transmitted signals' PDF and structured sparsity exhibited improved performance compared to \textbf{Scheme 1a)}. Specifically, it achieves a BER of $10^{-3}$ at 14.85 dB of SNR, which is 3.16 dB lower than that of \textbf{Scheme 1a)}, but 10.36 dB higher than the SNR of \textbf{Scheme 1g)}. The EP algorithm of \textbf{Scheme 1c)} imposes a high computational complexity due to iterative updates of the messages and requires a careful selection of approximations for the posterior distributions. However, it substantially improves the performance of the MU detection scheme.  As shown in Fig.~\ref{Figure:s1}, \textbf{Scheme 1c)} achieves a BER of $10^{-4}$ at an SNR of 10 dB. The conventional DNN detector is trained by the data of ML detection and it requires about 8.3 dB SNR for attaining a BER of $10^{-4}$.\par

The GNN-aided detector of \textbf{Scheme 1d)}-\textbf{Scheme 1f)} exhibits a performance gain by relying on an accurate PDF of the transmitted signal information. Firstly, \textbf{Scheme 1d)} iteratively updates the estimated signal with the aid of the MMSE \textit{a posteriori} information and achieves a BER of $10^{-4}$ at 12.22 dB of SNR. Then, for the GNN-AMP detector of \textbf{Scheme 1e)},  the simple Gaussian distribution approximation of the transmitted signal distribution associated with the iteratively updated GNN module also leads to substantial performance improvement compared to \textbf{Scheme 1b)}, which achieves a BER of $10^{-4}$ at 7.37 dB. This is only 0.3 dB  worse than the SNR of \textbf{Scheme 1g)}, desprite its significant reduction in complexity. Furthermore, \textbf{Scheme 1f)} shows near-ML performance albeit at an increased detection complexity at a BER of $10^{-4}$.

As the APM order increases from $\mathcal{L}=2$ to $\mathcal{L}=4$, the BER performance of all detection schemes deteriorates due to the reduced minimum Euclidean distance between constellation points, making the system more susceptible to noise and interference. As shown in Fig.~\ref{Figure:s2}, \textbf{Scheme 2h)} achieves the best performance, reaching a BER of $10^{-4}$ at an SNR of approximately 7.02 dB. \textbf{Scheme 2a)} demonstrates significantly worse performance, requiring around 16.32 dB to achieve the BER of $10^{-2}$. This is due to its limited ability to exploit signal sparsity or \textit{a priori} information. \textbf{Scheme 2b)}  fails to reach the BER of $10^{-2}$ within the SNR range considered, indicating its weak performance. By contrast, \textbf{Scheme 2c)} can still achieve a BER of $10^{-4}$ at 10.13 dB, outperforming both the MMSE and AMP detectors, thanks to its use of local Gaussian approximations and iterative updates, albeit  at a higher computational cost.

The DNN-based detector of \textbf{Scheme 2d)} is trained with the aid of ML based detection and reaches a BER of $10^{-4}$ at 14.78 dB, showing about 10 dB gap compared to the ML benchmark. This gap indicates a limitation in the generalization capacity of the DNN model in multi-user  environments. On the other hand, the GNN-aided detectors of \textbf{Scheme 2e)} and \textbf{Scheme 2f)} significantly improve the performance due to their ability to incorporate graph-based message passing that models the inter-user interference structure.

Specifically, \textbf{Scheme 2e)}  achieves a BER of $10^{-4}$ at an SNR of 12.21 dB, while \textbf{Scheme 2f)}  reaches the same BER at 7.37 dB. This is only 0.35 dB worse than \textbf{Scheme 2h)}, making it an efficient design alternative, as a benefit of its  dramatically reduced complexity. \textbf{Scheme 2g)}, which amalgamates a graph-based structure and expectation propagation principles, further improves the performance and achieves a BER of $10^{-4}$ at 7.09 dB, approaching the optimal ML performance with the aid of a practical detection architecture.

Fig.~\ref{Figure:s3} illustrates the performance of \textbf{Scheme 3} for $U=16$ users, 4 TAs per user and 64 RAs deployed. As the number of users increases, the ML complexity  increases exponentially, which becomes excessive. Compared to Fig.~\ref{Figure:s1}, \textbf{Scheme 3a)} has worse performance due to the increased number of users.  As shown in Fig.~\ref{Figure:s3}, \textbf{Scheme 3b)} achieves similar performance to \textbf{Scheme 3a)}, failing to achieve a BER of $10^{-3}$ even at 25 dB SNR. Then, \textbf{Scheme 3c)} has even worse performance in this situation, exhibiting a BER of $10^{-4}$ at 24.82 dB SNR.\par

Next, the GNN-based detector of \textbf{Scheme 3d)}-\textbf{Scheme 3f)} is characterized in Fig.~\ref{Figure:s3}. \textbf{Scheme 3d)} does not perform well in this case, resulting in similar performance to \textbf{Scheme 3a)} and \textbf{Scheme 3b)}. \textbf{Scheme 3e)} shows a good performance in $U=16$ users achieving a BER of $10^{-4}$ at 20.16 dB SNR, while \textbf{Scheme 3f)} performs 1.8 dB better than the \textbf{Scheme 3e)}.

In \textbf{Scheme 4}, we double the number of RAs to attain additional diversity gain. As shown in Fig.~\ref{Figure:s4}, the performance of \textbf{Scheme 4a)} and \textbf{Scheme 4b)} slightly improves. \textbf{Scheme 4c)} can succeeds in exploiting the benefits of increased number of RAs, then it achieves a BER of $10^{-4}$ at 18.65 dB SNR, which is about 6 dB better the performance of \textbf{Scheme 3}.  With an increased number of RAs, all GNN-aided detectors of \textbf{Scheme 4} achieve improved performance compared to \textbf{Scheme 3}.  The \textbf{Scheme 4d)} detector obtains a BER of $10^{-3}$ at 21.8 dB SNR, while \textbf{Scheme 3d)} attains a BER of $10^{-2}$ at 24.37 dB SNR. Both \textbf{Scheme 4e)} and \textbf{Scheme 4f)} achieve a BER of $10^{-3}$ at about 15.52 dB, where \textbf{Scheme 4e)} only suffers from a 0.2 dB SNR degradation with respect to \textbf{Scheme 4f)}.

\begin{figure}[!htb]
  \centering
  \includegraphics[width=8.8cm]{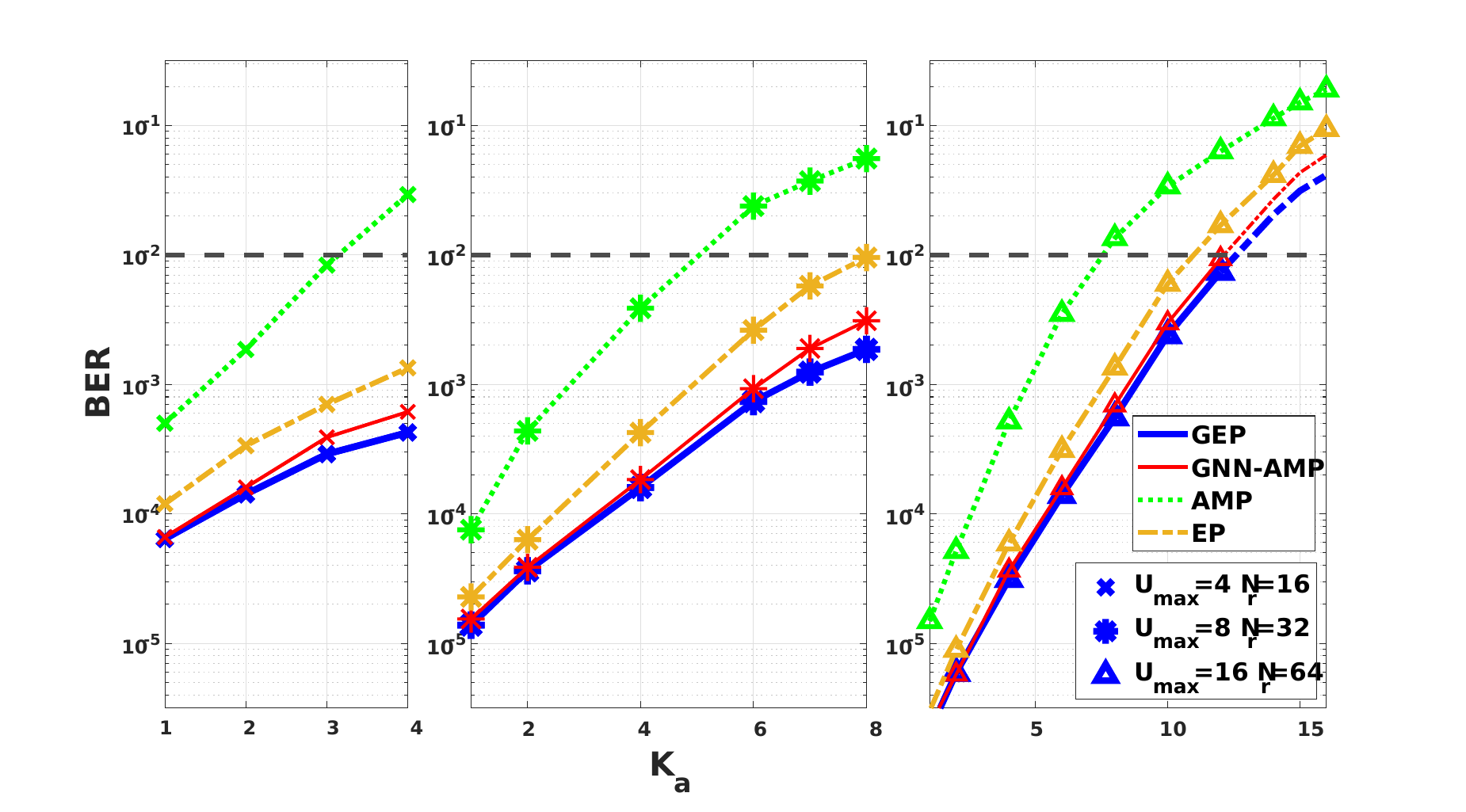}
 \caption{BER performance versus the number of active users $K_a$ for $N_r=16,32,64.$ at SNR of 5 dB for transmission over Rayleigh channel.}
  \label{Figure:varying}
\end{figure}

Additionally, we  consider a more realistic case, where not all users are are active simultaneously. Fig.~\ref{Figure:varying} illustrates the detection performance of our MU CS-SFIM framework  under different maximum numbers of users, $U_{\max} = {4, 8, 16}$, in a (MU-UL setting. In order to compare the robustness of both GNN-aided methods, the GEP and GNN-AMP with conventional MP methods, namely AMP and EP under a fixed SNR of 5 dB while varying the number of active users $K_a$. A threshold of BER at $10^{-2}$ is used as the benchmark for effective transmission. Each curve, distinguished by different markers,  corresponds to a distinct scheme  evaluated for scenarios with $K_a$ ranging from 1 up to the maximum number of users. All GNN-aided methods are trained by a scheme that applies a mixture of a single active user to the maximum number of users to see the full range, ensuring fair comparison across different loads.\par
Across all scenarios, both GNN-AMP and GEP demonstrate significantly better BER performance than conventional AMP and EP, particularly as the number of active users increases. For instance, when  $U_{max}=4$, AMP suffers from a rapid BER degradation exceeding a BER of  $10^{-2}$ even at a low load of $K_a=3$, while the GNN-aided methods and EP still maintain reasonable performance under $10^{-2}$ of BER. For a $U_{max}=8$ scheme, the AMP method reaches the  $10^{-2}$ of BER at $K_a=5$, while EP reaches the  $10^{-2}$ of BER when all users are activated. GNN-AMP and GEP maintain viable performance under the same settings.  In the scenario of a larger maximum user load of $U_{max}=16$ and large scale of RAs, each scheme can achieves better performance compared with less $U_{max}$ and less RAs. However, with the load of user increasing, detection performance of each scheme degraded and exceed the effective threshold at a BER of  $10^{-2}$.  The GNN solutions maintain a viable BER of $10^{-2}$ at $K_a=12$, while the AMP method is ineffective at $K_a=8$ and EP exceeds a BER of $10^{-2}$ at $K_a=11$. The EP methods can outperform the AMP based low-complexity detector under varying of active users' but at a high computational complexity. Harnessing a GNN in both the AMP and EP can further increase the detection performance, despite varying the number of users.

Fig.~\ref{Figure:varying}also shows that the GEP provides a marginally better performance than GNN-AMP, albeit at a higher complexity. GNN-AMP, on the other hand, achieves a better trade-off between complexity and performance. Furthermore, while all algorithms suffer from a degraded performance upon increasing $K_a$, the GNN-aided detectors exhibit a marginally better performance in MU-UL environments than conventional MP-based detectors.

\begin{figure}[!htb]
  \centering
  \includegraphics[width=8.8cm]{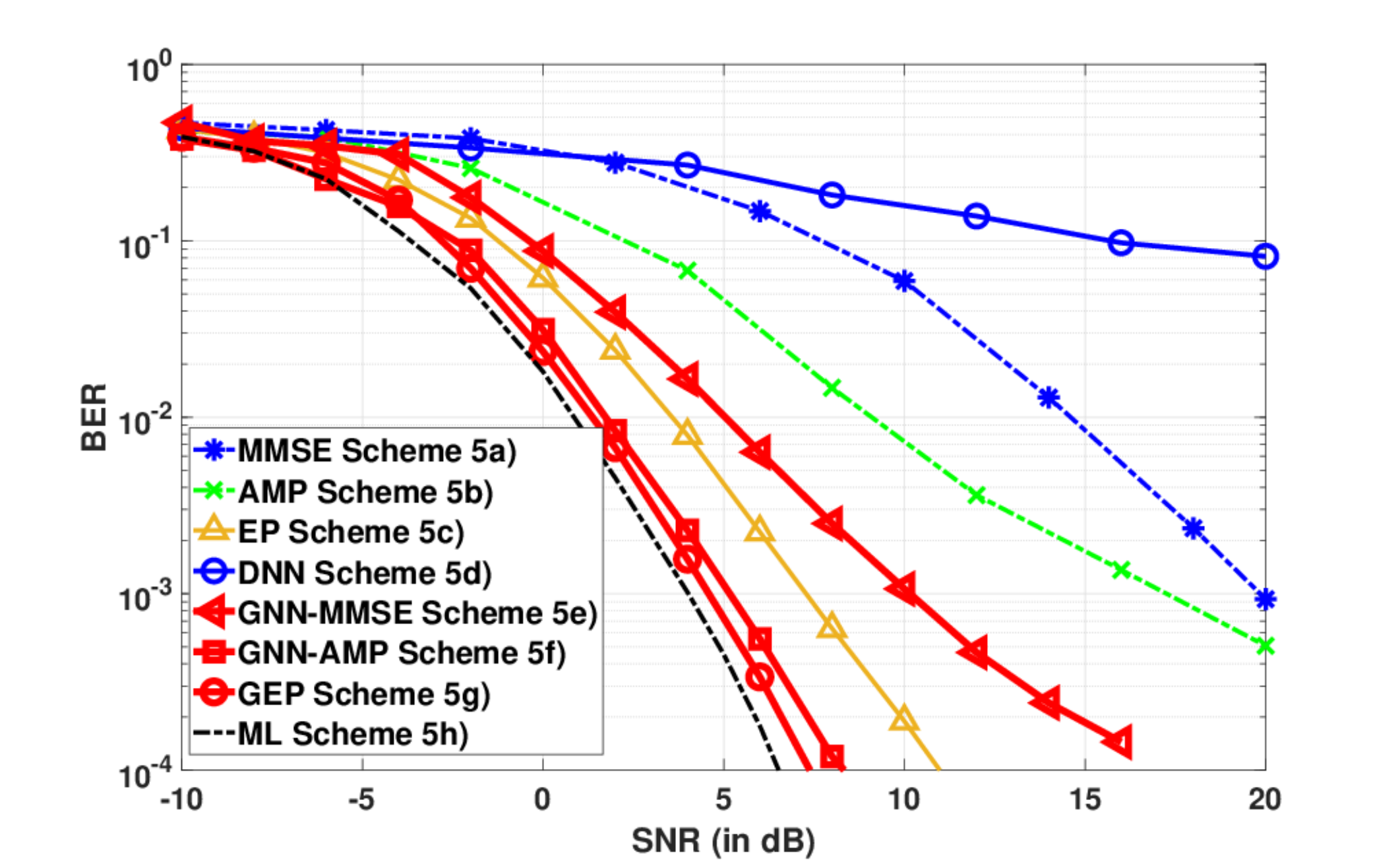}
 \caption{Performance comparison of detectors in Scheme 5 for transmission over Rayleigh channel.}
  \label{Figure:s5}
\end{figure}

 For example, the \textbf{Scheme 5} settings are used for characterizing the performance, where we consider $K_a=3$, 3 out of 4 users being activated at a time. The DNN-based detector is trained with a fixed number of antennas and users. However, the number of users in practical MIMO systems changes dynamically.  As shown in Fig.~\ref{Figure:s5}, we train the GNN-aided methods for $U=4$ users and employ a $16\times16$ MIMO size. Then, we simulate the  scenario, where 3 of the 4 users are communicating simultaneously. As shown in Fig.~\ref{Figure:s5}, both \textbf{Scheme 5a)} and \textbf{Scheme 5b)} result in performance degradation  with respect to  \textbf{Scheme 1}. As the MP methods have the ability to deal with a dynamic number of users, \textbf{Scheme 5c)} still performs well in this case, as shown in Fig.~\ref{Figure:s5}.  Furthermore, both \textbf{Scheme 5e)} and \textbf{Scheme 5f)} exhibit robustness to a fluctuating number of users, due to the flexibility of the GNN.

\section{CONCLUSION}
\label{sec:conc}
GNN-based detectors were conceived for MU-CS-SFIM  systems, namely the GNN-MMSE, AMP-GNN and GEPnet by providing the GNN module with the accurate distribution of \textit{a priori} information and by intrinsically amalgamating the GNN module with MP algorithms. The MP-based GNN methods benefit from the low complexity of the MP algorithm and the improved efficiency brought by the GNN module.  We demonstrated  that the GNN based MP algorithms significantly outperform the corresponding conventional MP methods. The simulation results also confirmed that the GEPnet outperformed AMP-GNN at the cost of a higher complexity.  Furthermore, we also validated that the GNN-aided detectors, particularly GNN-AMP and GEPnet, exhibit robustness to dynamic variations in the number of active users, which demonstrates that the GNN-based detectors are capable of supporting a flexible number of users. Furthermore, since the \textit{a priori} information has already been exploited by our MP-based GNN detector, channel coding can be naturally incorporated into the proposed GNN-based framework for iteratively exchanging soft information by harnessing joint detection and decoding. This constitutes an interesting area for future research.

\bibliographystyle{IEEEtran}
\bibliography{main}

\begin{thebibliography}{10}
\providecommand{\url}[1]{#1}
\csname url@samestyle\endcsname
\providecommand{\newblock}{\relax}
\providecommand{\bibinfo}[2]{#2}
\providecommand{\BIBentrySTDinterwordspacing}{\spaceskip=0pt\relax}
\providecommand{\BIBentryALTinterwordstretchfactor}{4}
\providecommand{\BIBentryALTinterwordspacing}{\spaceskip=\fontdimen2\font plus
\BIBentryALTinterwordstretchfactor\fontdimen3\font minus
  \fontdimen4\font\relax}
\providecommand{\BIBforeignlanguage}[2]{{%
\expandafter\ifx\csname l@#1\endcsname\relax
\typeout{** WARNING: IEEEtran.bst: No hyphenation pattern has been}%
\typeout{** loaded for the language `#1'. Using the pattern for}%
\typeout{** the default language instead.}%
\else
\language=\csname l@#1\endcsname
\fi
#2}}
\providecommand{\BIBdecl}{\relax}
\BIBdecl

\bibitem{mutut}
M.~Jiang and L.~Hanzo, ``{Multiuser MIMO-OFDM for Next-Generation Wireless
  Systems},'' \emph{Proceedings of the IEEE}, vol.~95, no.~7, pp. 1430--1469,
  2007.

\bibitem{MIMOdet}
S.~Yang and L.~Hanzo, ``{Fifty Years of MIMO Detection: The Road to Large-Scale
  MIMOs},'' \emph{IEEE Communications Surveys and Tutorials}, vol.~17, no.~4,
  pp. 1941--1988, 2015.

\bibitem{massivemimo1}
M.~A. Albreem, M.~Juntti, and S.~Shahabuddin, ``{Massive MIMO Detection
  Techniques: A Survey},'' \emph{IEEE Communications Surveys and Tutorials},
  vol.~21, no.~4, pp. 3109--3132, 2019.

\bibitem{massivemimo2}
K.~K. Vaigandla and D.~N. Venu, ``{Survey on Massive MIMO: Technology,
  Challenges, Opportunities and Benefits},'' 2021.

\bibitem{im1}
Y.~Chau and S.-H. Yu, ``{Space Modulation On Wireless Fading Channels},'' in
  \emph{{IEEE 54th Vehicular Technology Conference. VTC Fall 2001. Proceedings
  (Cat. No.01CH37211)}}, vol.~3, 2001, pp. 1668--1671 vol.3.

\bibitem{smhistory}
N.~Ishikawa, S.~Sugiura, and L.~Hanzo, ``{50 Years of Permutation, Spatial and
  Index Modulation: From Classic RF to Visible Light Communications and Data
  Storage},'' \emph{IEEE Communications Surveys and Tutorials}, vol.~20, no.~3,
  pp. 1905--1938, 2018.

\bibitem{sm1}
R.~Y. Mesleh, H.~Haas, S.~Sinanovic, C.~W. Ahn, and S.~Yun, ``{Spatial
  Modulation},'' \emph{IEEE Transactions on Vehicular Technology}, vol.~57,
  no.~4, pp. 2228--2241, 2008.

\bibitem{chaosm}
C.~Xu, S.~Sugiura, S.~X. Ng, and L.~Hanzo, ``{Spatial Modulation and Space-Time
  Shift Keying: Optimal Performance at a Reduced Detection Complexity},''
  \emph{IEEE Transactions on Communications}, vol.~61, no.~1, pp. 206--216,
  2013.

\bibitem{massivesm}
L.~He, J.~Wang, and J.~Song, ``{On Massive Spatial Modulation MIMO: Spectral
  Efficiency Analysis and Optimal System Design},'' in \emph{2016 IEEE Global
  Communications Conference (GLOBECOM)}, 2016, pp. 1--6.

\bibitem{smdesign}
P.~Yang, M.~Di~Renzo, Y.~Xiao, S.~Li, and L.~Hanzo, ``{Design Guidelines for
  Spatial Modulation},'' \emph{IEEE Communications Surveys and Tutorials},
  vol.~17, no.~1, pp. 6--26, 2015.

\bibitem{gsm}
T.~Datta and A.~Chockalingam, ``{On Generalized Spatial Modulation},'' in
  \emph{{2013 IEEE Wireless Communications and Networking Conference (WCNC)}},
  2013, pp. 2716--2721.

\bibitem{gsmmu}
T.~Lakshmi~Narasimhan, P.~Raviteja, and A.~Chockalingam, ``{Generalized Spatial
  Modulation in Large-Scale Multiuser MIMO Systems},'' \emph{IEEE Transactions
  on Wireless Communications}, vol.~14, no.~7, pp. 3764--3779, 2015.

\bibitem{imconcept}
E.~Basar, ``{Index Modulation Techniques For 5G Wireless Networks},''
  \emph{IEEE Communications Magazine}, vol.~54, no.~7, pp. 168--175, 7 2016.

\bibitem{sim}
R.~Abu-alhiga and H.~Haas, ``{Subcarrier-Index Modulation OFDM},'' \emph{2009
  IEEE 20th International Symposium on Personal, Indoor and Mobile Radio
  Communications}, pp. 177--181, 2009.

\bibitem{ofdmim1}
M.~Wen, X.~Cheng, M.~Ma, B.~Jiao, and H.~V. Poor, ``{On the Achievable Rate of
  OFDM With Index Modulation},'' \emph{IEEE Transactions on Signal Processing},
  vol.~64, no.~8, pp. 1919--1932, 2016.

\bibitem{gofdmim}
R.~Fan, Y.~J. Yu, and Y.~L. Guan, ``{Generalization of Orthogonal Frequency
  Division Multiplexing With Index Modulation},'' \emph{IEEE Transactions on
  Wireless Communications}, vol.~14, no.~10, pp. 5350--5359, 2015.

\bibitem{muofdmim}
M.~Yüzgeçcioğlu and E.~Jorswieck, ``{Uplink and Downlink Transceiver Design
  for OFDM with Index Modulation in Multi-user Networks},'' in \emph{{2017 IEEE
  28th Annual International Symposium on Personal, Indoor, and Mobile Radio
  Communications (PIMRC)}}, 2017, pp. 1--5.

\bibitem{csofdmim}
H.~Zhang, L.-L. Yang, and L.~Hanzo, ``{Compressed Sensing Improves the
  Performance of Subcarrier Index-Modulation-Assisted OFDM},'' \emph{IEEE
  Access}, vol.~4, pp. 7859--7873, 2016.

\bibitem{cs}
D.~L. Donoho, ``{Compressed Sensing},'' \emph{IEEE Transactions on Information
  Theory}, vol.~52, no.~4, pp. 1289--1306, apr 2006.

\bibitem{sfim}
T.~Datta, H.~S. Eshwaraiah, A.~Chockalingam, and S.~Member, ``{Generalized
  Space-and-Frequency Index Modulation},'' \emph{IEEE Transactions on Vehicular
  Technology}, vol.~65, no.~7, pp. 4911--4924, 2016.

\bibitem{mflsm}
I.~A. Hemadeh, M.~El-Hajjar, and L.~Hanzo, ``{Hierarchical Multi-Functional
  Layered Spatial Modulation},'' \emph{IEEE Access}, vol.~6, pp. 9492--9533,
  2018.

\bibitem{mim1}
B.~Shamasundar, S.~Bhat, S.~Jacob, and A.~Chockalingam, ``{Multidimensional
  Index Modulation in Wireless Communications},'' \emph{IEEE Access}, vol.~6,
  pp. 589--604, 2018.

\bibitem{mim2}
P.~Yang, Y.~Xiao, Y.~L. Guan, M.~Di~Renzo, S.~Li, and L.~Hanzo, ``{Multidomain
  Index Modulation for Vehicular and Railway Communications: A Survey of Novel
  Techniques},'' \emph{IEEE Vehicular Technology Magazine}, vol.~13, no.~3, pp.
  124--134, 2018.

\bibitem{chaomim}
C.~Xu, Y.~Xiong, N.~Ishikawa, R.~Rajashekar, S.~Sugiura, Z.~Wang, S.-X. Ng,
  L.-L. Yang, and L.~Hanzo, ``{Space-, Time- and Frequency-Domain Index
  Modulation for Next-Generation Wireless: A Unified Single-/Multi-Carrier and
  Single-/Multi-RF MIMO Framework},'' \emph{IEEE Transactions on Wireless
  Communications}, vol.~20, no.~6, pp. 3847--3864, 2021.

\bibitem{csmim}
S.~Lu, I.~A. Hemadeh, M.~El-Hajjar, and L.~Hanzo, ``{Compressed Sensing-Aided
  Multi-Dimensional Index Modulation},'' \emph{IEEE Transactions on
  Communications}, vol.~67, no.~6, pp. 4074--4087, 2019.

\bibitem{jmim}
X.~Feng, M.~El-Hajjar, C.~Xu, and L.~Hanzo, ``{Near-Instantaneously Adaptive
  Learning-Assisted and Compressed Sensing-Aided Joint Multi-Dimensional Index
  Modulation},'' \emph{IEEE Open Journal of Vehicular Technology}, vol.~4, pp.
  893--912, 2023.

\bibitem{mimmu}
S.~Lu, M.~El-Hajjar, and L.~Hanzo, ``{Two-Dimensional Index Modulation for the
  Large-Scale Multi-User MIMO Uplink},'' \emph{IEEE Transactions on Vehicular
  Technology}, vol.~68, no.~8, pp. 7904--7918, 2019.

\bibitem{mpcs}
D.~L. Donoho, A.~Maleki, and A.~Montanari, ``{Message Passing Algorithms for
  Compressed Sensing},'' \emph{Proceedings of the National Academy of
  Sciences}, vol. 106, no.~45, pp. 18\,914--18\,919, 2009.

\bibitem{mimoamp}
C.~Jeon, R.~Ghods, A.~Maleki, and C.~Studer, ``{Optimality of Large MIMO
  Detection via Approximate Message Passing},'' in \emph{{2015 IEEE
  International Symposium on Information Theory (ISIT)}}, 2015, pp. 1227--1231.

\bibitem{ampgsmofdmim}
L.~Wei, J.~Zheng, and Q.~Liu, ``{Approximate Message Passing Detector for Index
  Modulation With Multiple Active Resources},'' \emph{IEEE Transactions on
  Vehicular Technology}, vol.~68, no.~1, pp. 972--976, 2019.

\bibitem{ampofdmim}
Z.~Sui, S.~Yan, H.~Zhang, L.-L. Yang, and L.~Hanzo, ``{Approximate Message
  Passing Algorithms for Low Complexity OFDM-IM Detection},'' \emph{IEEE
  Transactions on Vehicular Technology}, vol.~70, no.~9, pp. 9607--9612, 2021.

\bibitem{gsfimamp}
B.~Chakrapani, T.~L. Narasimhan, and A.~Chockalingam, ``{Generalized
  Space-Frequency Index Modulation: Low-Complexity Encoding and Detection},''
  in \emph{{2015 IEEE Globecom Workshops (GC Wkshps)}}, 2015, pp. 1--6.

\bibitem{oamp}
J.~Ma and L.~Ping, ``{Orthogonal AMP},'' \emph{IEEE Access}, vol.~5, pp.
  2020--2033, 2017.

\bibitem{ep}
J.~Céspedes, P.~M. Olmos, M.~Sánchez-Fernández, and F.~Perez-Cruz,
  ``{Expectation Propagation Detection for High-Order High-Dimensional MIMO
  Systems},'' \emph{IEEE Transactions on Communications}, vol.~62, no.~8, pp.
  2840--2849, 2014.

\bibitem{epmimo}
H.~Wang, A.~Kosasih, C.-K. Wen, S.~Jin, and W.~Hardjawana, ``{Expectation
  Propagation Detector for Extra-Large Scale Massive MIMO},'' \emph{IEEE
  Transactions on Wireless Communications}, vol.~19, no.~3, pp. 2036--2051,
  2020.

\bibitem{learnmimo}
H.~He, C.-K. Wen, S.~Jin, and G.~Y. Li, ``{Model-Driven Deep Learning for MIMO
  Detection},'' \emph{IEEE Transactions on Signal Processing}, vol.~68, pp.
  1702--1715, 2020.

\bibitem{dlphy}
H.~He, S.~Jin, C.-K. Wen, F.~Gao, G.~Y. Li, and Z.~Xu, ``{Model-Driven Deep
  Learning for Physical Layer Communications},'' \emph{IEEE Wireless
  Communications}, vol.~26, no.~5, pp. 77--83, 2019.

\bibitem{dldet}
N.~Samuel and T.~Diskin, ``{Learning to Detect},'' \emph{IEEE Transactions on
  Signal Processing}, vol.~67, no.~10, pp. 2554--2564, may 2019.

\bibitem{gnnmimo}
A.~Scotti, N.~N. Moghadam, D.~Liu, K.~Gafvert, and J.~Huang, ``{Graph Neural
  Networks for Massive MIMO Detection},'' \emph{arXiv preprint
  arXiv:2007.05703}, 2020.

\bibitem{gnnmu}
Y.~Shen, J.~Zhang, S.~H. Song, and K.~B. Letaief, ``{Graph Neural Networks for
  Wireless Communications: From Theory to Practice},'' \emph{IEEE Transactions
  on Wireless Communications}, vol.~22, no.~5, pp. 3554--3569, 2023.

\bibitem{gnnampmumimo}
H.~He, A.~Kosasih, X.~Yu, J.~Zhang, S.~Song, W.~Hardjawana, and K.~B. Letaief,
  ``{GNN-Enhanced Approximate Message Passing for Massive/Ultra-Massive MIMO
  Detection},'' in \emph{{2023 IEEE Wireless Communications and Networking
  Conference (WCNC)}}, 2023, pp. 1--6.

\bibitem{gnnepmimo}
A.~Kosasih, V.~Onasis, W.~Hardjawana, V.~Miloslavskaya, V.~Andrean, J.-S. Leu,
  and B.~Vucetic, ``{Graph Neural Network Aided Expectation Propagation
  Detector for MU-MIMO Systems},'' in \emph{{2022 IEEE Wireless Communications
  and Networking Conference (WCNC)}}, 2022, pp. 1212--1217.

\bibitem{gnnmumimo}
A.~Kosasih, V.~Onasis, V.~Miloslavskaya, W.~Hardjawana, V.~Andrean, and
  B.~Vucetic, ``{Graph Neural Network Aided MU-MIMO Detectors},'' \emph{IEEE
  Journal on Selected Areas in Communications}, vol.~40, no.~9, pp. 2540--2555,
  2022.

\bibitem{immu}
J.~Li, Q.~Li, S.~Dang, M.~Wen, X.-Q. Jiang, and Y.~Peng, ``{Low-Complexity
  Detection for Index Modulation Multiple Access},'' \emph{IEEE Wireless
  Communications Letters}, vol.~9, no.~7, pp. 943--947, 2020.

\bibitem{musm}
S.~Narayanan, M.~J. Chaudhry, A.~Stavridis, M.~Di~Renzo, F.~Graziosi, and
  H.~Haas, ``{Multi-User Spatial Modulation MIMO},'' in \emph{{2014 IEEE
  Wireless Communications and Networking Conference (WCNC)}}, 2014, pp.
  671--676.

\bibitem{epsoftmu}
K.-H. Ngo, M.~Guillaud, A.~Decurninge, S.~Yang, S.~Sarkar, and P.~Schniter,
  ``{Non-Coherent Multi-User Detection Based on Expectation Propagation},'' in
  \emph{{2019 53rd Asilomar Conference on Signals, Systems, and Computers}},
  2019, pp. 2092--2096.

\bibitem{ismofdm}
Y.~Liu, M.~Zhang, H.~Wang, and X.~Cheng, ``{Spatial Modulation Orthogonal
  Frequency Division Multiplexing with Subcarrier Index Modulation for V2X
  Communications},'' in \emph{{2016 International Conference on Computing,
  Networking and Communications (ICNC)}}, 2016, pp. 1--5.

\bibitem{ZhangCompressedOFDM}
H.~Zhang, L.-L. Yang, and L.~Hanzo, ``{Compressed Sensing Improves the
  Performance of Subcarrier Index-Modulation-Assisted OFDM},'' \emph{IEEE
  Access}, vol.~4, pp. 7859--7873, 2016.

\bibitem{mumimoofdm}
M.~Jiang and L.~Hanzo, ``{Multiuser MIMO-OFDM for Next-Generation Wireless
  Systems},'' \emph{Proceedings of the IEEE}, vol.~95, no.~7, pp. 1430--1469,
  2007.

\bibitem{munonlinear}
S.~Chen, S.~X. Ng, E.~F. Khalaf, A.~Morfeq, and N.~D. Alotaibi, ``{Multiuser
  Detection for Nonlinear MIMO Uplink},'' \emph{IEEE Transactions on
  Communications}, vol.~68, no.~1, pp. 207--219, 2020.

\bibitem{musyn1}
E.~Ekrem, M.~Koca, and H.~Deliç, ``{Iterative Synchronization of Multiuser
  Ultra-Wideband Signals},'' \emph{IEEE Transactions on Wireless
  Communications}, vol.~9, no.~10, pp. 3040--3051, 2010.

\bibitem{musyn2}
W.~M. Jang, B.~Vojcic, and R.~Pickholtz, ``{Joint transmitter-receiver
  optimization in synchronous multiuser communications over multipath
  channels},'' \emph{IEEE Transactions on Communications}, vol.~46, no.~2, pp.
  269--278, 1998.

\bibitem{sumprod}
F.~Kschischang, B.~Frey, and H.-A. Loeliger, ``{Factor Graphs and the
  Sum-Product Algorithm},'' \emph{IEEE Transactions on Information Theory},
  vol.~47, no.~2, pp. 498--519, 2001.

\bibitem{ofdmimamp}
Z.~Sui, S.~Yan, H.~Zhang, L.-L. Yang, and L.~Hanzo, ``{Approximate Message
  Passing Algorithms for Low Complexity OFDM-IM Detection},'' \emph{IEEE
  Transactions on Vehicular Technology}, vol.~70, no.~9, pp. 9607--9612, 2021.

\bibitem{smamp}
L.~Wei, J.~Zheng, and Q.~Liu, ``{Approximate Message Passing Detector for Index
  Modulation With Multiple Active Resources},'' \emph{IEEE Transactions on
  Vehicular Technology}, vol.~68, no.~1, pp. 972--976, 2019.

\bibitem{Gaussianapp}
J.~P. Vila and P.~Schniter, ``{Expectation-Maximization Gaussian-Mixture
  Approximate Message Passing},'' \emph{IEEE Transactions on Signal
  Processing}, vol.~61, no.~19, pp. 4658--4672, 2013.

\bibitem{ampsimp}
S.~Rangan, ``{Generalized Approximate Message Passing for Estimation with
  Random Linear Mixing},'' in \emph{{2011 IEEE International Symposium on
  Information Theory Proceedings}}, 2011, pp. 2168--2172.

\bibitem{epbayesian}
T.~P. Minka, ``{Expectation Propagation for Approximate Bayesian Inference},''
  \emph{arXiv preprint arXiv:1301.2294}, 2013.

\bibitem{gnnreview}
B.~Khemani, S.~Patil, K.~Kotecha, and S.~Tanwar, ``{A review of graph neural
  networks: concepts, architectures, techniques, challenges, datasets,
  applications, and future directions},'' \emph{Journal of Big Data}, vol.~11,
  no.~1, p.~18, 2024.

\bibitem{gru}
K.~Cho, B.~Van~Merri{\"e}nboer, C.~Gulcehre, D.~Bahdanau, F.~Bougares,
  H.~Schwenk, and Y.~Bengio, ``{Learning Phrase Representations using RNN
  Encoder-Decoder for Statistical Machine Translation},'' \emph{arXiv preprint
  arXiv:1406.1078}, 2014.

\bibitem{xinyudnn}
X.~Feng, M.~El-Hajjar, C.~Xu, and L.~Hanzo, ``Near-instantaneously adaptive
  learning-assisted and compressed sensing-aided joint multi-dimensional index
  modulation,'' \emph{IEEE Open Journal of Vehicular Technology}, vol.~4, pp.
  893--912, 2023.

\end{thebibliography}

\end{document}